\documentclass[aps,pre,reprint,superscriptaddress,longbibliography,floatfix]{revtex4-1}

\usepackage{graphicx}
\usepackage{bm}
\usepackage{color}
\usepackage{amsmath}
\usepackage{amssymb}
\usepackage{dsfont}
\usepackage{color}
\usepackage{calc}
\usepackage{hyperref}
\usepackage{multirow}
\usepackage{natmove}
\usepackage{placeins} 

\usepackage{grffile}
\usepackage[caption=false]{subfig}

\newcommand{\lla}{\left\langle}
\newcommand{\rra}{\right\rangle}

\graphicspath{{./figures/}}

\begin{document}
\title{Clustering of microswimmers: Interplay of shape and hydrodynamics}

\author{Mario Theers}
\affiliation{Theoretical Soft Matter and Biophysics, Institute for
Advanced Simulation and Institute of Complex Systems,
Forschungszentrum J\"ulich, D-52425 J\"ulich, Germany}
\author{Elmar Westphal}
\affiliation{Peter Gr\"unberg Institute and J\"ulich Centre for Neutron
Science, Forschungszentrum J\"ulich, D-52425 J\"ulich, Germany}
\author{Kai Qi}
\affiliation{Theoretical Soft Matter and Biophysics, Institute for
Advanced Simulation and Institute of Complex Systems,
Forschungszentrum J\"ulich, D-52425 J\"ulich, Germany}
\author{Roland G. Winkler}
\email{r.winkler@fz-juelich.de}
\affiliation{Theoretical Soft Matter and Biophysics, Institute for
Advanced Simulation and Institute of Complex Systems,
Forschungszentrum J\"ulich, D-52425 J\"ulich, Germany}
\author{Gerhard Gompper}
\email{g.gompper@fz-juelich.de}
\affiliation{Theoretical Soft Matter and Biophysics, Institute for
Advanced Simulation and Institute of Complex Systems,
Forschungszentrum J\"ulich, D-52425 J\"ulich, Germany}

\date{\today}

\begin{abstract}

The spatiotemporal dynamics in systems of active self-propelled particles is controlled by
the propulsion mechanism in combination with various direct interactions, such as steric repulsion,
hydrodynamics, and chemical fields. Yet, these direct interactions are typically anisotropic,
and come in different ``flavors'', such as spherical and elongated particle shapes for steric
repulsion, pusher and puller flow fields for hydrodynamics, etc. The combination of the various aspects is expected to lead to new emergent behavior.  However, it is a priori not evident whether shape and hydrodynamics act synergistically or antagonistically to generate motility-induced clustering (MIC) and phase separation (MIPS).
We employ a model of prolate spheroidal microswimmers---called
squirmers---in quasi-two-dimensional confinement to address this issue by mesoscale
hydrodynamic simulations. For comparison, non-hydrodynamic active Brownian particles (ABPs)
are considered to elucidate the contribution of hydrodynamic interactions on MIC and MIPS.
For spherical particles, the comparison between ABP and hydrodynamic-squirmer ensembles reveals
a suppression of MIPS due to hydrodynamic interactions. Yet, our analysis shows that
dynamic clusters exist, with a broad size distribution. The fundamental difference
between ABPs and squirmers is attributed to
an increased reorientation of squirmers by hydrodynamic torques during their collisions.
In contrast, for elongated squirmers, hydrodynamics interactions enhance MIPS. The transition
to a phase-separated state strongly depends on the nature of the swimmer's flow field
---with an increased tendency toward MIPS for pullers, a reduced tendency for pushers.
Thus, hydrodynamic interactions show opposing effects on MIPS for spherical and elongated
microswimmers. Our results imply that details of the propulsion mechanism of biological
microswimmers, like pattern and time dependence of the flagellar beat, may be very important
to determine their collective behavior.


\end{abstract}
\maketitle

\section{Introduction}

Motile bacteria at interfaces exhibit intriguing collective phenomena \cite{elge:15,marc:13}, such as cluster formation, observed for {\em Myxococcus
xanthus} \cite{peru:12} or  {\em Thiovulum majus} \cite{petr:15}, as well as swarming, swirling, raft formation  \cite{kear:10,darn:10,wu:00,domb:04,cope:09,zhan:10}, and the emergence of mesoscale turbulence \cite{mend:99,domb:04,wens:12,wens:12.1}, observed for {\em E. coli}. Similarly, experiments on self-phoretic artificial spherical microswimmers, such as Janus particles, self-propelled liquid droplets, and photo-activated colloids, exhibit cluster formation and phase separation despite their isotropic shape and purely repulsive interactions \cite{gole:05,theu:12,thut:11,pala:13,butt:13,marc:16.1,bech:16}. The various active agents are propelled by different mechanisms and may exhibit different steric and propulsion-related  interactions. Hence, it is a priori not evident which processes govern structure formation in the rather distinct systems. To unravel the underlaying universal features and to discriminate them from specific aspects requires dedicated experimental and theoretical studies.

Computer simulations of model systems, such as spherical active Brownian particles (ABPs), yield motility-induced clustering (MIC) and motility-induced phase separation (MIPS) \cite{fily:12,redn:13,bial:12,marc:16.1,wyso:14,sten:14,wyso:16} in qualitative agreement with  the above-mentioned experiments on synthetic self-phoretic particles.
The intuitive explanation for the emergence of MIC and MIPS is a positive feedback between blocking of persistent particle motion by steric interactions, and an enhanced probability of collisions with further particles at sufficiently large concentrations and activities \cite{bech:16,mata:14,cate:15}. However,
in self-phoretic systems, the phoretic field has been show to significantly contribute to clustering by an field-induced  attractive interaction \cite{gole:12,pohl:14}.


Cluster formation, swarming, and mesoscale turbulence of anisotropic objects such as bacteria is explained by steric interaction-induced alignment. Computer simulations of two-dimensional assemblies of rods suggest that the interplay of rod geometry, self-propulsion, and steric interaction suffices to facilitate aggregation into clusters \cite{peru:06,peru:12,yang:10,abke:13} and even lead to swarming and turbulence \cite{wens:12,wens:12.1}.

Aside from steric, other interaction mechanisms between active particles are present, evidently and most prominently fluid-mediated interactions, since motile bacteria and phoretic microswimmers propel themselves via the embedding fluid. Hence, an crucial aspect regarding MIC and MIPS of microswimmers is the role of hydrodynamics. At equilibrium, hydrodynamic interactions (HIs) solely affect the dynamics and not the structural properties of a system. Since active systems are intrinsically out of equilibrium, this does not necessary apply, and the steady-state---hence the formation of clusters---is strongly affected by hydrodynamics \cite{zoet:14,mata:14,delm:15,alar:17,yosh:17}.

The phase behavior of active particles in the presence of hydrodynamic interactions has received much less attention than ABPs. In simulations, this is largely due to the substantial computational challenges posed by hydrodynamics. Specifically, the long-ranged flow field created by the swimmers has to be accounted for adequately. In experiments, HI is difficult to switch off, and therefore its contribution hard to assess. In contrast, simulations can be performed with and without HI and the latter effect on MIPS be analysed, although with an increased numerical effort.

A frequently applied model for a self-propelled particles in the presence of a fluid is a squirmer \cite{ligh:52,blak:71,ishi:06.1,pago:13,goet:10,zoet:14,thee:16.1}---a spheroidal colloid with a prescribed slip velocity on its surface. It was originally introduced to describe ciliated microswimmers such as \textit{Paramecia} and {\em Volvox}. Nowadays, it is applied to a broad class of microswimmers, both, biological as well as synthetic ones.  A squirmer is typically characterized by two modes accounting for its swimming velocity and its active stress. The latter distinguishes between pushers (e.g., {\em E. coli}, sperm), pullers (e.g., {\em Chlamydomonas}), and neutral squirmers (e.g., {\em Paramecium}) \cite{thee:16.1}.

In order to shed light onto the importance of hydrodynamics for structure formation in active systems, we perform mesoscale hydrodynamic simulations applying the multiparticle collision dynamics approach (MPC), a particle-based hydrodynamic simulation method to solve the Navier-Stokes equations \cite{male:99,kapr:08,gomp:09}. Specifically, we consider squirmers confined in a thin slit, i.e., in a quasi-2D geometry. Here, we want to address and clarify several issues.   On the one hand, we want to resolve the contradiction between Refs.~\cite{zoet:14} and \cite{mata:14}, which predict enhanced clustering of spherical squirmers or no MIPS in the presence of HIs, respectively. On the other hand, we consider squirmers of prolate spheroidal shape. This is motivated by the different mechanism of cluster formation for spherical and elongated particles (blocking vs. alignment). Thus, the effect of hydrodynamic interactions on the collective behavior of spherical swimmers can be qualitatively different than for swimmers with elongated shapes.

As an important result, we find no MIPS for spherical squirmers in quasi-2D, in agreement with Ref.~\cite{mata:14} and \cite{yosh:17}, but in contrast to Ref.~\cite{zoet:14}. We attribute the discrepancy to peculiarities of the compressible MPC fluid employed in Ref.~\cite{zoet:14}, which requires a suitable choice of the  parameters for simulations of low-compressibility fluids. Hence, HIs suppress cluster formation and MIPS of self-propelled spherical squirmers compared to ABPs. Interestingly, the opposite is true for spheroidal squirmers with sufficiently large aspect ratio, where we find a substantial enhancement of MIPS by HIs. This applies for pushers, pullers, and neutral squirmers, as long as the force dipole is sufficiently weak. In case of large force dipoles, pushers show the least tendency to phase separation, with even ABPs exhibiting stronger cluster formation. A density-aspect-ratio phase diagram for moderate force dipole strength shows most pronounced phase separation for pullers, followed by neutral squirmers, pushers, and finally ABPs. Thus, the effects of shape and hydrodynamics are highly non-additive, and act sometimes synergistically, sometimes antagonistically.

The paper is structured as follows. Previous results of squirmer simulations of collective phenomena  are briefly summarized in Sec.~\ref{sec:previous_sim}.
Section~\ref{sec:MPC} introduces the model for prolate squirmers.
A comparison between MPC and ABP simulation results of spherical squirmers is presented in Sec. \ref{sec:results_spherical}. Section~\ref{sec:results_spheroidal} presents results for prolate spheroidal swimmers, and Sec.~\ref{sec:discussion} discusses the various obtained aspects.   Section \ref{sec:conclusion} summarizes our results.
Appendix~\ref{app:simulation}  presents the  simulation approach for the fluid (MPC)  (\ref{app:mpc}), the coupling of a squirmer with the fluid (\ref{app:bc}), as well as  the  simulation method for ABPs (\ref{app:abp}).
Appendix~\ref{app:SpheroidalABP} collects definitions relevant for rigid-body dynamic of ABPs, and  App.~\ref{app:inhomogeneities} discusses technical aspects of fluid compressibility in particle-based simulations of squirmers.


\section{Collective Behavior of Squirmers---Brief Summary of Previous Simulation Studies} \label{sec:previous_sim}

The dynamics of spheroidal squirmers in two dimensions (2D) and in a three-dimensional slit geometry (quasi-2D) has been studied and compared with ABPs in Refs.~\cite{mata:14,yosh:17} and \cite{zoet:14,blas:16}, respectively. In Ref.~\cite{mata:14},  the squirmer dynamics is strictly two dimensional, but embedded in a 3D fluid. The shape of the squirmers is changed from infinitely long cylinders (perpendicular to the 2D plane), corresponding to 2D HIs, to highly flattened cylinders with 3D HIs.
Independent of the cylinder length (and dimensionality of HIs), the simulation studies show no evidence of  phase separation, in contrast to comparable simulations of ABPs. The qualitatively different behavior is attributed to a faster decorrelation of the squirmer's swimming direction due to HIs compared to the configuration-independent rotational diffusion of ABPs. The studies of Ref.~\cite{yosh:17} emphasizes the importance of near-field HIs in the suppression of MIPS. However, clustering of the squirmers is observed and the formation of polar order for nearly neutral squirmers. Note that in both Refs.~\cite{mata:14,yosh:17}, thermal fluctuations are neglected. Hence,  there is no thermal rotational diffusion leading to a decorrelation of the squirmers orientational motion.
In the simulations of Refs.~\cite{zoet:14,blas:16}, squirmers are confined in a narrow slit (quasi-2D), slightly wider then their diameter, which  captures the geometry of experiments, where glass plates are used for confinement, and thermal fluctuations are taken into account. Here, the swimming direction is free to orient in three dimensions, and hydrodynamics is screened and decays as $1/\varrho^2$ parallel to the confining surfaces in the slit center, where $\varrho$ is the radial distance from the squirmer \cite{math:16.1}. Most importantly, hydrodynamics is argued to enhances MIPS, in contrast to findings of Refs.~\cite{mata:14,yosh:17}, and a phase diagram is presented in Ref.~\cite{blas:16}.

A quasi-2D setup is certainly less favorable for cluster formation of ABPs compared to a strict  2D system,  since the swimmer's propulsion direction can point toward a wall, which reduces steric blocking and the lateral swim pressure \cite{taka:14,taka:15.1,wink:15}. Thus, compared to strictly 2D ABP systems, ABPs in quasi-2D require larger  P\'eclet numbers and packing fractions to form clusters, and those clusters  exhibit a finite life time. Surface hydrodynamic interactions may lead to preferred squirmer orientations. Here, considerations based on far-field HIs predict a parallel alignment of the propulsion direction for pushers and a normal orientation for pullers at a no-slip surface \cite{spag:12}. Hence, quasi-2D confinement may result in a different behavior, and near-field effects (specifically with the confining walls) might be important for cluster formation as indicated by the studies of Refs.~\cite{zoet:14,thee:16.1}.

Suspensions of athermal spherical squirmers in 3D, lattice-Boltzmann \cite{alar:13} and force-coupling \cite{delm:15} simulations have also found a pronounced polar order for pullers and neutral squirmers  and clustering over a certain range of force-dipole strengths. However, no large-scale MIPS is observed. In contrast, ABPs in 3D clearly exhibit MIPS \cite{wyso:14,sten:14,wyso:16} and large-scale collective motion \cite{wyso:14,wyso:14}. Moreover, simulations of 2D, quasi-2D, and 3D systems of attractive squirmers show a substantially enhanced cluster formation compared to purely repulsive squirmers \cite{alar:17,delm:15}. Also in such situations, hydrodynamics is found to reduce MIPS compared to ABPs.

So far, the influence of HIs on MIPS has mainly been studied for spherical squirmers. In contrast, very little is known about the influence of HIs on the structure and dynamics of elongated spheroidal squirmers confined in narrow slits. Here, the shape, squirmer-squirmer HIs, and HIs with the confining surfaces  affect cluster formation and a possible MIC and  MIPS \cite{thee:16.1}.

\section{Model and Simulation Approach} \label{sec:MPC}


\begin{figure}[t]
\begin{center}
\includegraphics*[width=\columnwidth]{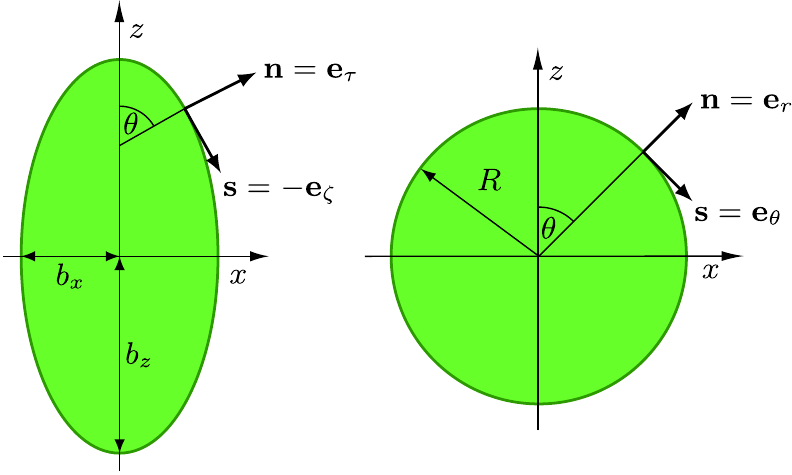}
\caption{\label{fig:sketch_spheroid}
  Illustration of normal and tangent vectors of a spheroidal (left) and spherical (right) squirmer. Self-propulsion along the body-fixed orientation vector $\bm e$, here $\bm e = (0,0,1)^T$, is achieved by an axisymmetric prescribed surface velocity in the direction of the tangent vector $\bm{s}$  \cite{thee:16.1}.
}
\end{center}
\end{figure}

We model a nonspherical squirmer as a prolate spheroidal rigid body with the prescribed surface velocity (cf. Fig. \ref{fig:sketch_spheroid}) \cite{kell:77,lesh:07,ishi:13,thee:16.1}
\begin{align} \label{Eq:Def_usq_of_zeta}
   \bm{u}_{sq}=-B_1 (\bm{e}_\zeta \cdot \bm{e}_z) (1+\beta \zeta) \bm{e}_\zeta .
\end{align}
Here, we employ spheroidal coordinates $(\zeta, \tau, \varphi)$ (cf. Fig.~\ref{fig:sketch_spheroid}), which are related to Cartesian coordinates via
\begin{align} \label{Eq:xyz_of_tau_zeta_phi} \nonumber
  x &= c \sqrt{\tau^2-1}\sqrt{1-\zeta^2} \cos \varphi , \\
  y &= c \sqrt{\tau^2-1}\sqrt{1-\zeta^2} \sin \varphi , \\ \nonumber
  z&=c \tau \zeta ,
\end{align}
where $c=\sqrt{b_z^2-b_x^2}$, $-1 \leq \zeta \leq 1$, $1 \leq \tau < \infty$, and $0 \leq \varphi \leq 2 \pi$,  which is an extension of  and alternative to earlier approaches \cite{kell:77,ishi:13}. The advantage of the choice (\ref{Eq:Def_usq_of_zeta}) for the slip velocity is that an analytical solution of the flow field can be obtained  \cite{thee:16.1}. The constant $B_1$ in Eq.~(\ref{Eq:Def_usq_of_zeta}) determines the amplitude of the slip velocity and is related to the self-propulsion velocity $U_0$ via \cite{thee:16.1}
\begin{align} \label{eq:u0}
  U_0=B_1 \tau_0 \left( \tau_0-(\tau_0^2-1)\coth^{-1} \tau_0 \right) ,
\end{align}
where $\tau_0=b_z/c$.  The active stress is captured by the coefficient $\beta$ \cite{ishi:06,thee:16.1}. Thereby, $\beta < 0$ corresponds to a pusher, $\beta > 0$ to a puller, and $\beta=0$ to a neutral squirmer.

\begin{figure}
\includegraphics*[width=\columnwidth]{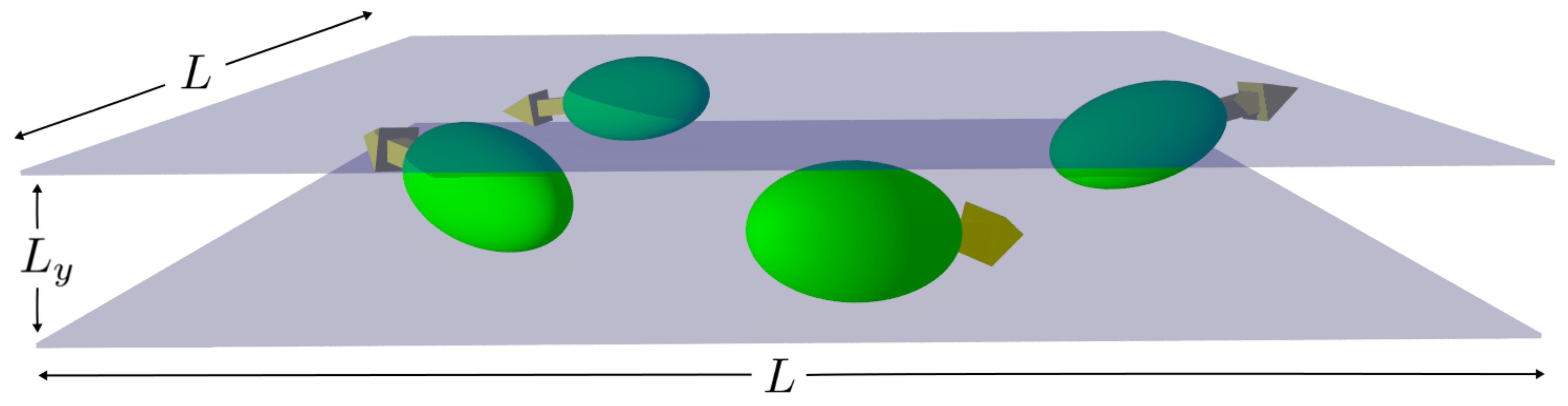}
\caption{\label{fig:slit_sketch}
  Sketch of spheroidal microswimmers in a  narrow  slit. The ratio of the squirmer minor axis and the width of the slit is $2b_x/L_y = 6/7$. The arrows indicate the swimming direction along a squirmers major axis.
}
\end{figure}

We consider the collective dynamics of squirmers confined in a narrow slit of dimensions $L \times L_y \times L$ (cf. Fig.~\ref{fig:slit_sketch}), where the ratio of the squirmer minor axis and the width of the slit is  $2b_x/L_y=6/7$, i.e., we focus on a quasi-2D geometry.

The squirmers are embedded in a fluid, which we simulate by multiparticle collision dynamics (MPC), a mesoscale hydrodynamics simulation approach  \cite{male:99,kapr:08,gomp:09}. The details of the algorithm are described in App.~\ref{app:mpc}, and the implementation of the squirmer and the coupling with MPC  in App.~\ref{app:bc}.

For comparison, we perform simulations of active Brownian spheres and spheroids. The respective simulation approach is described in App.~\ref{app:abp}.

\section{Simulations---Spherical Squirmers} \label{sec:results_spherical}

\subsection{System Setup and Parameters} \label{sec:setup_low}

We consider two setups for the studies of spherical swimmers, with different packing fractions and propulsion velocities. Common to both cases is the number of swimmers, $N_{sw} = 196$, the radius of the squirmer, $R=3a$, the with of the slit, $L_y=7a$, and the active-stress parameter, $\beta \in \{ -1,0,1\}$.\\

\paragraph*{Set 1 -- High packing fraction, small P\'eclet number:}
The choice of  the box length  $L=96a = 32 R$ corresponds to a 3D packing fraction of $\phi=4\pi N_{sw} R^3/ 3 L^2 L_y = 0.34$, or the quasi-2D packing fraction
\begin{align}
 \phi^{2D}= \frac{N_{sw} \pi R^2}{L^2}=0.6.
\end{align}
We employ the time step $h=0.02 \sqrt{ma^2/(k_BT)}$ and the mean number of particles in a cell $\lla N_c \rra= 80$, which yields a fluid viscosity of $\eta=178 \sqrt{m k_B T/a^4}$, as determined by independent simulations \cite{thee:15}.
The resulting P\'eclet number, which compares the time scale for rotational diffusion, $2D_R$ ($D_R$ is the rotational diffusion coefficient),  to the swimming time scale, $2R/U_0$, is
\begin{align}
  Pe= \frac{U_0}{2R D_R} =115
\end{align}
for the swim velocity $U_0 = 2B_1/3$, with  $B_1=0.01 \sqrt{k_BT/m}$, and $D_R \approx 10^{-5} \sqrt{k_BT/m a^2}$ from simulations. The latter is approximately $20\%$ larger than the theoretical value.
The chosen value of $\lla N_c\rra=80$  is large compared to typical MPC studies, and results in an increased computational effort.
However, it leads to a high viscosity, hence, a low $D_R$ and high P\'eclet number, a prerequisite to observe cluster formation in ABPs, as well as a low fluid compressibility. In Ref. \cite{zoet:12}, a high P\'eclet number was achieved by a large $B_1$ and hence $U_0$.
However, we find that such high values of $B_1$ lead to fluid-density inhomogeneities and consequently simulation artifacts (see App.~\ref{app:inhomogeneities}).
Complementary to the squirmer simulations with MPC, we perform ABP simulations with the same values for $U_0$, $D_R$, $R$, and $L_y$, but consider additionally the  larger system size  $L=192a$ in order to reduce finite-size effects.

\begin{figure}[t]
  \centering
    \includegraphics[width=0.9\columnwidth]{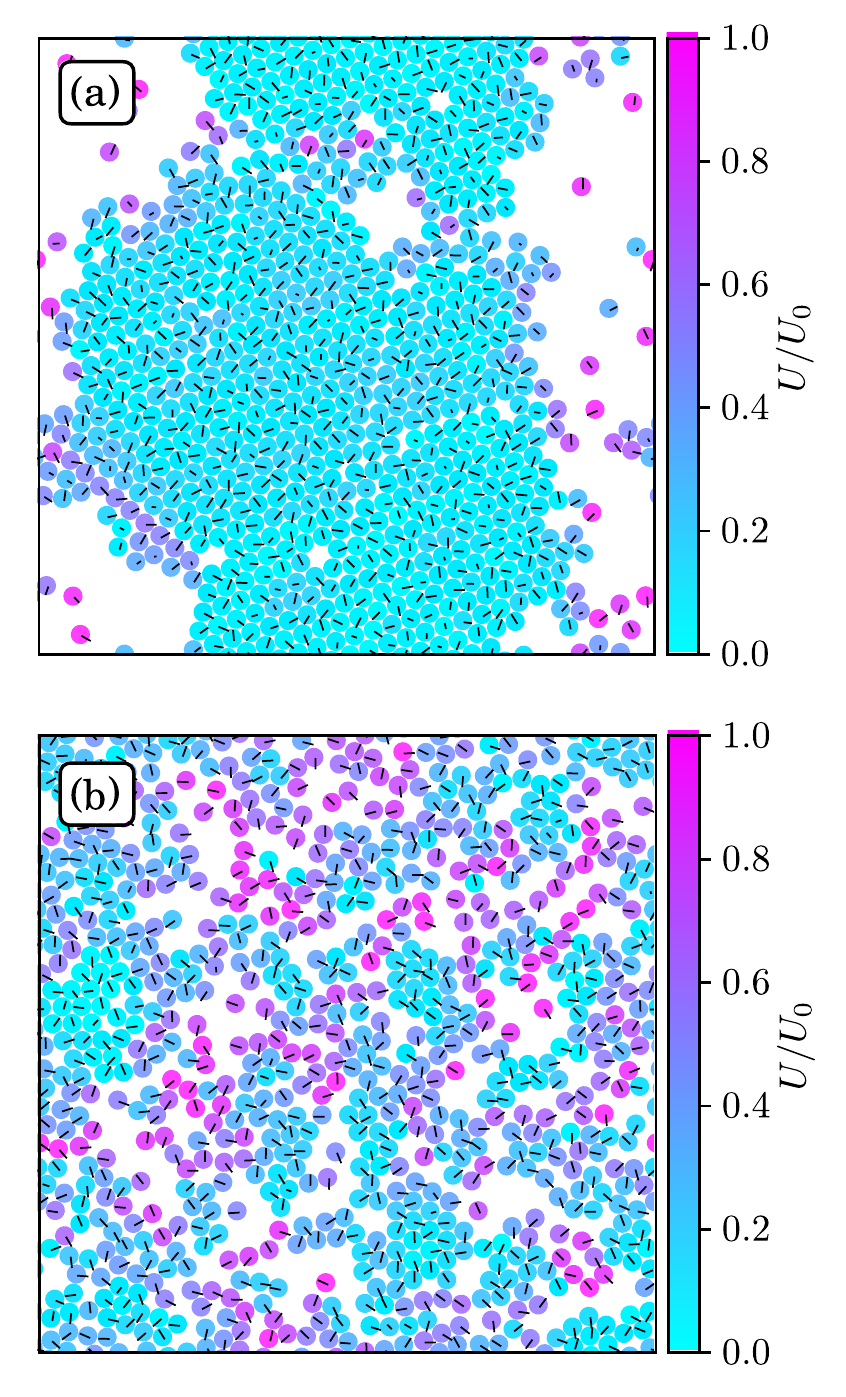}
  \caption{\label{fig:snapshots_RHO80}
  Snapshots of spherical squirmers at $Pe=115$, the packing fraction $\phi^{2D}=0.6$, and the simulation box length $L=192a$. (a) Active Brownian particles exhibit MIPS and local hexagonal order. (b) Neutral squirmers ($\beta=0$) exhibit no long-range order and no MIPS. Note that the cluster in (a) partly  dissolves in the course of time and a new  system-spanning cluster forms subsequently. Similarly, the clusters in (b) are unstable. See Supplemental Material at ??? for movies of the two cases (M1, M2). }
\end{figure}

\paragraph*{Set 2 -- Various packing fractions, large P\'eclet number:}
The box lengths $L/a=96$, $106$, and $119$ yield the packing fractions
$\phi^{2D}=0.6$, $0.5$, and $0.4$, respectively.
The MPC parameters $h=0.05 \sqrt{ma^2/k_BT}$ and $\lla N_c \rra =480$ result in a high viscosity $\eta=445 \sqrt{m k_B T/a^4}$ and, with $B_1 = 0.02 \sqrt{k_B T/m}$, the large P\'eclet number
$Pe=575$.

\begin{figure}[t]
  \centering
    \includegraphics[width=\columnwidth]{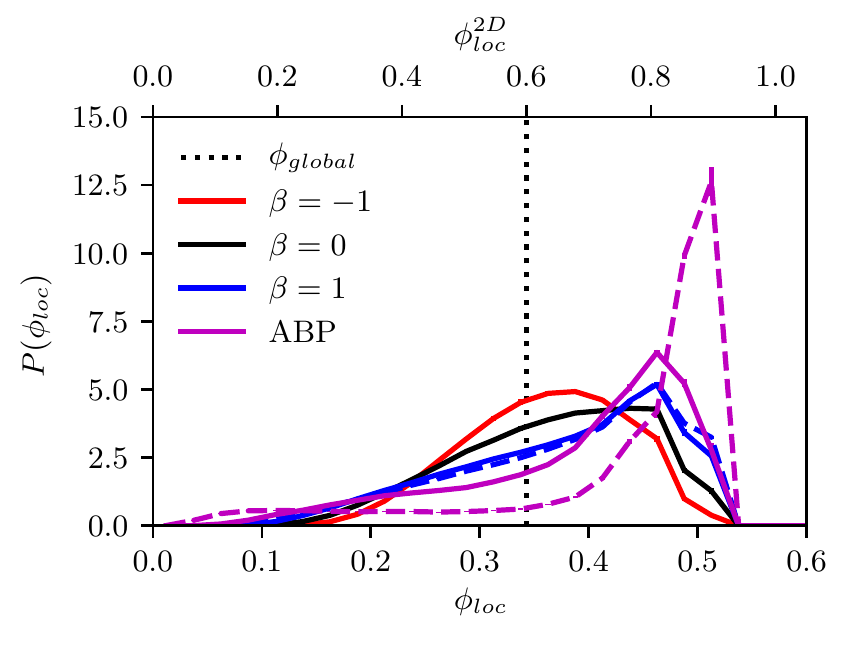} \\
    \includegraphics[width=\columnwidth]{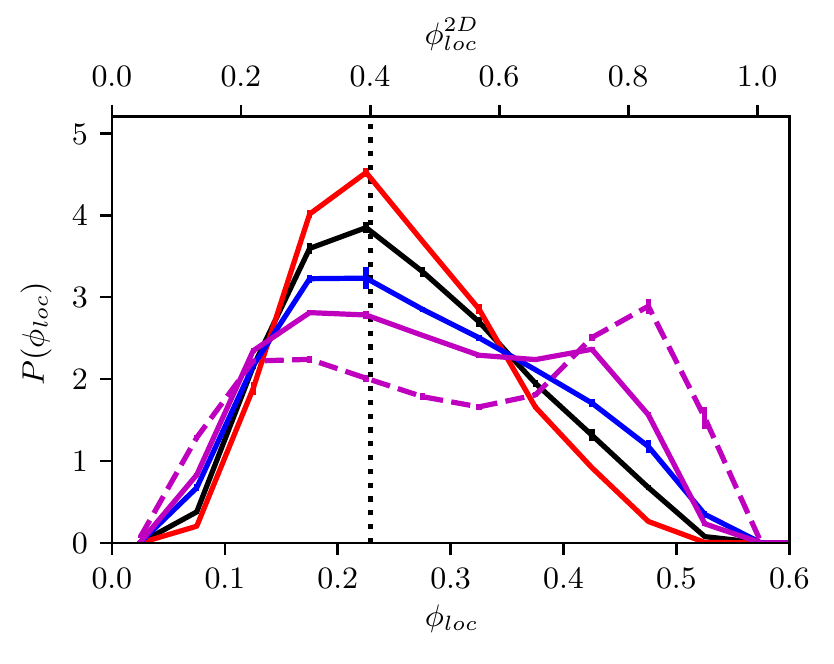}
  \caption{Probability distribution of local packing fractions  of spherical squirmers and ABPs (purple) for (a)  the P\'eclet number $Pe=115$ and the global area packing fraction $\phi^{2D}=0.6$, and (b) $Pe=575$ and  $\phi^{2D}=0.4$. Results for pushers ($\beta=-1$, red), pullers ($\beta =1$, blue), and neutral squirmers ($\beta =0$, black) are displayed. The solid and dashed lines correspond to the system sizes $L=96a$ and $L=192a$, respectively.   \label{fig:MIPS_suppressed_spherical}}
\end{figure}

\subsection{Results}

Figure~\ref{fig:snapshots_RHO80} displays snapshots of structures of spherical ABPs and squirmers for the parameter {\em Set 1}. The ABP system exhibits strong clustering and hexagonal order, whereas no large clusters and no pronounced order emerges for neutral squirmers. Hence, hydrodynamics strongly affects the aggregation behavior.

 In order to characterize the clustering behavior more quantitatively, we analyse the density distribution by performing a Voronoi decomposition of the fluid volume with the squirmer's centers as generator points \cite{rycr:09,wyso:14}. A local packing fraction $\phi_{loc}$ of squirmers is then introduced as the ratio of a squirmer volume to the volume of its Voronoi cell. Note that the Voroni construction yields in general a distribution function which is different from the distribution function of the local density. In dilute regions, $\phi_{loc}$ is low, since the Voronoi volume is large, and $P(\phi_{loc})$ is low too, because of the low particle density.

\begin{figure}
\centering
\includegraphics*[width=\columnwidth]{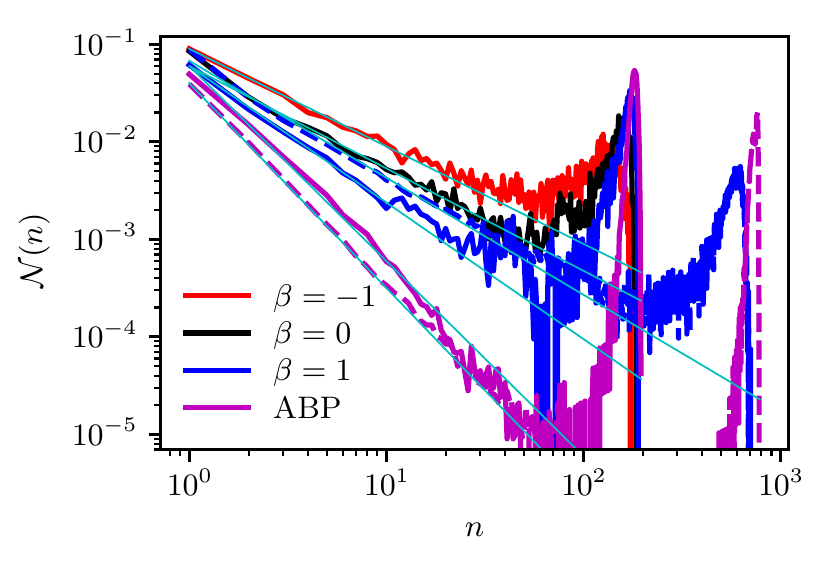}
\caption{\label{fig:cluster_size}
Cluster-size distribution function for squirmers (pusher: $\beta=-1$, puller: $\beta =1$, neutral: $\beta =0$) and ABPs for the packing fraction $\phi^{2D}=0.6$ and the P\'eclet number $Pe=575$. The solid and dashed lines correspond to the system sizes $L=96a$ and $L=192a$, respectively. The lines (light blue) indicate the power-law decay $P(n) \sim n^{-\delta}$, with $\delta = 1$ (pusher: $\beta=-1$), $1$ (neutral: $\beta =0$), $1.4$ (puller: $\beta =1$), and $\delta =2$, $2.1$ for the small and large ABP system, respectively.
}
\end{figure}

Figure~\ref{fig:MIPS_suppressed_spherical}(a) shows  distributions of local packing fractions for squirmers and ABPs at $Pe=115$.  We consider two system sizes for the ABPs and pullers ($\beta=1$), namely $L=96 a$ and $L=192a$, to provide clear evidence of a possible MIPS. The ABP system exhibits pronounced peaks at the packing fractions $\phi^{2D}_{loc} \approx 0.8$ ($L/a=96$, solid) and $\phi^{2D}_{loc} \approx 0.85$ ($L/a=192$, dashed), respectively, close to the maximum value of $\phi^{2D}=0.907$ of 2D hexagonal packing,  and lower probabilities for dilute regions. The two curves reflect a strong system-size dependence, as is well-known for first-order phase transitions, because the system is affected by interfaces \cite{chal:86}. Accordingly, and in agreement with previous simulations \cite{bial:12,fily:12,redn:13}, our APB system phase separates into a dense giant cluster comprising the vast majority of the ABPs (cf. Fig.~\ref{fig:cluster_size}) and a dilute gas-like region. In contrast, squirmers show a far less pronounced high-density region. Most remarkable, the distribution function $P(\phi_{loc})$ is independent of the system size, as indicated in Fig.~\ref{fig:MIPS_suppressed_spherical}(a) for pullers, which we consider as indication that squirmers exhibit no MIPS. The force dipole evidently strongly affects cluster formation as is reflected by the differences of the distribution functions. Thereby, pulling ($\beta =1$) promotes formation of larger  clusters, whereas pushing ($\beta =-1$)is  disadvantage for cluster formation. At the lower packing fraction $\phi^{2D} = 0.4$ and the high P\'eclet number $Pe= 575$ (Fig.~\ref{fig:MIPS_suppressed_spherical}(b)), ABPs exhibit a two-peak distribution, which becomes more pronounced with increasing system size, and peaks shift to smaller and larger concentrations, respectively, as expected for a first order phase transition, indicating a MIPS in the limit $L\to \infty$. For the squirmers, the maximum of the local density for pullers and neutral squirmers coincides with the global density $\phi^{2D}=0.4$. However, for pullers a long tail toward higher local densities indicates formation of large temporary clusters (MIC). Considering the higher packing fraction $\phi^{2D}=0.6$ for $Pe=575$, we find little differences to the density distributions of Fig.~\ref{fig:MIPS_suppressed_spherical}(a).

Another quantity to characterize the structure is the  cluster-size distribution function ${\cal N}(n)$, which is defined  as
\begin{align} \label{eq:cluster_distribution}
{\cal N}(n) = \frac{1}{N_{sw}}  n p(n) ,
\end{align}
where ${\cal N}(n)$ is the average fraction of particles that are members
of a cluster of size $n$, and  $p(n)$ is  the number of clusters
of size $n$. The distribution is normalized such that $\sum_{n=1}^{N_{sw}} {\cal N}(n) =1$. We use a distance criterion to define a cluster: a swimmer belongs to a cluster, when its center-to-center distance to another swimmer of the cluster is within $1.02 \sigma$.  As shown in Fig.~\ref{fig:cluster_size}, predominately smaller clusters are formed for squirmers. However, ABPs exhibit a peak for clusters comprising almost all swimmers, consisted with MIPS. This is particularly evident for the larger system size.  With increasing system size, the probability for intermediate-size clusters drops significantly and ${\cal N}(n)$ is dominated by small and giant clusters. However, the initial power-law decay of ${\cal N}(n)$ changes only slightly from ${\cal N}(n)_{ABP} \sim n^{-2.0}$ for $L=96a$ to ${\cal N}(n)_{ABP} \sim n^{-2.1}$ for $L=192a$. For larger systems, we expect an even more pronounced  giant cluster consistent with the appearance of MIPS.  Compared to ABPs, squirmers exhibit a higher probability for medium-size clusters in the range $2 < n \lesssim 150$.
The cluster-size distributions decay as a power-law  for $n \lesssim 30$. Thereby, pusher exhibit the slowest decay with ${\cal N}(n) \sim n^{-1.0}$,  and pullers  the somewhat fastest decay  ${\cal N}(n) \sim n^{-1.4}$. This power-law decrease of ${\cal N}(n)$ is independent of the system size.
The qualitative and quantitative behavior changes when the concentration is reduced (not shown). For $\phi^{2D}=0.4$, all systems exhibit an initial power-law decay ${\cal N}(n) \sim n^{-1}$  and no system-spanning cluster appears. However, pushers and neutral squirmers show a stronger, non-power-law decay for large cluster sizes. Hence, there is a crossover from a power-law to a faster decay of the cluster-size distribution function with decreasing concentration for $\beta \leqslant 0$. A similar behavior has been observed  in Ref.~\cite{alar:17}.

\begin{figure}
\centering
\includegraphics*[width=\columnwidth]{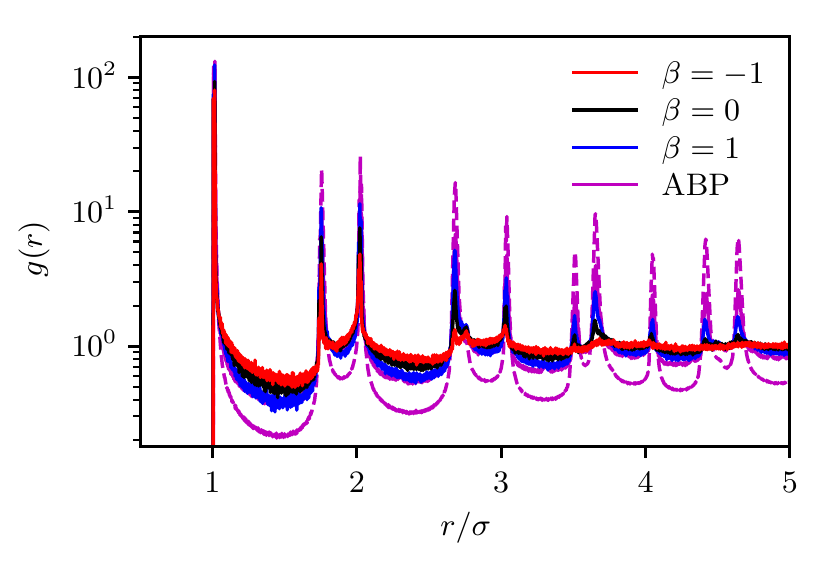}
\caption{\label{fig:pair_corr}
Two-dimensional pair-correlation functions for squirmers (pushers: $\beta=-1$, pullers: $\beta =1$, neutral squirmers: $\beta =0$) and ABPs for the packing fraction $\phi^{2D}=0.6$ and the P\'eclet number $Pe=575$. For ABPs, the system sizes are $L=96a$ (solid) and $L=192a$ (dashed).  Please note the logarithmic scale of the $y$-axis.
}
\end{figure}

 A pronounced long-range translational order for ABPs is clearly visible in the pair-correlation function $g(r)$ \cite{hans:86,alle:87} presented in Fig.~\ref{fig:pair_corr}. Even on the length scale of half of the system size $L$, peaks are visible for ABPs. Squirmers exhibit a less pronounced translation order for $\phi^{2D}=0.6$. Thereby, as already indicated in the local packing fraction, pullers ($\beta=1$) exhibit the most pronounced order, but the order decays faster with distance than for ABPs. Pushers ($\beta=-1$) exhibit the lowest tendency to long-rang order, and $g(r)$ assumes the ideal-gas value essentially on the scale of three to four neighbor distances.
A decreasing packing fraction leads to decreasing long-range order. For $\phi^{2D}=0.5$, APBs still show the most pronounced order, however, the correlation function reaches unity for $r/\sigma \gtrsim 7$. At $\phi^{2D}=0.4$, only short-range order is present for $r/\sigma \lesssim 4$. Interestingly, pullers exhibit the most pronounced order in this case.

\begin{figure}
  \centering
    \includegraphics[width=\columnwidth]{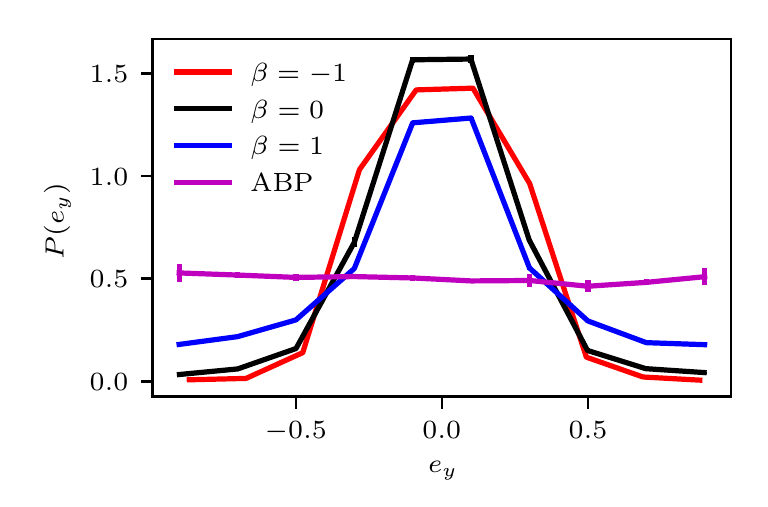}
  \caption{Probability distribution of the hexagonal order parameter $|q_6|^2$ of spherical ABPs (purple) and squirmers (pullers: $\beta=-1$, red, neutral: $\beta=0$, black, pullers: $\beta=1$, blue) for $Pe=115$ and $\phi^{2D}=0.6$. \label{fig:MIPS_suppressed_spherical_order}}
\end{figure}

The peak positions of Fig.~\ref{fig:pair_corr} are consistent with a hexagonal packing of the swimmers.
The preference of such an order is confirmed by the hexagonal order parameter $|q_6|^2$  defined as \cite{stei:83,bial:12,zoet:14}
\begin{align}
 q_6 \equiv \frac{1}{6} \sum_{j \in N_6^k} e^{i 6 \vartheta_{kj}},
\end{align}
where $N_6^k$ is the set of the six nearest neighbors of swimmer $k$, and $\vartheta_{kj}$ is the angle between the vector $\bm{R}_k-\bm{R}_j$ connecting the centers of particles $j$ and $k$ and the $x$-axis. For a perfect hexagonal lattice $|q_6|^2=1$.
Figure~\ref{fig:MIPS_suppressed_spherical_order} displays probability distribution functions of the order parameter for squirmers and ABPs at $Pe=115$. Distributions for the larger P\'eclet number $Pe=575$ closely agree with those of Fig.~\ref{fig:MIPS_suppressed_spherical_order}. Consistent with the high local packing, ABPs exhibit an pronounced probability for hexagonal order. The respective values for squirmers are smaller. Thereby, the tendency to form hexagonal order is most distinct for pullers. In contrast, pushers exhibit hardly any local hexagonal order.

\begin{figure}[t]
  \centering
    \includegraphics[width=\columnwidth]{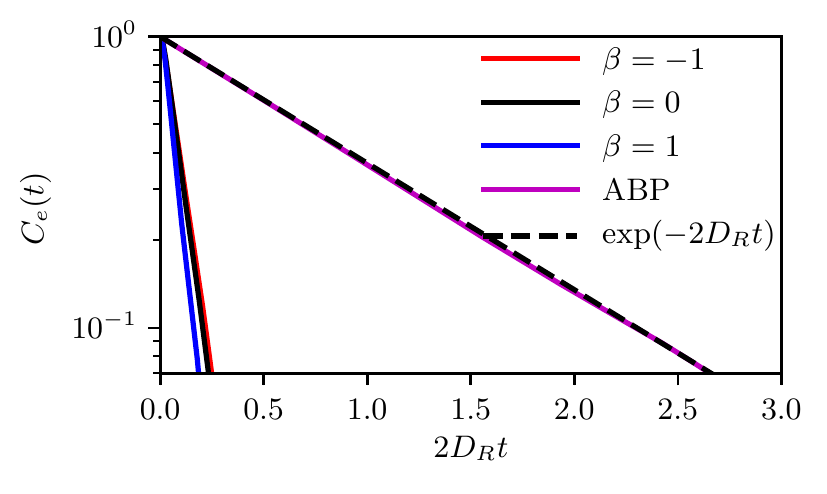}
  \caption{\label{fig:prop_orientation}
 Probability distribution functions of the propulsion-direction component $e_y$ normal to the confining walls for    ABPs and squirmers. The squirmers are preferentially orientation parallel to the walls. The P\'eclet number is  $Pe=575$ and the concentration $\phi^{2D}=0.6$. The distribution functions for $Pe=115$ and the same density are nearly identical.
  }
\end{figure}

Strictly 2D ABP systems exhibit stronger aggregation and a more pronounced phase separation \cite{zoet:14,bial:12} compared to similar systems in quasi-2D \cite{zoet:14}. The reason is that the freedom of the propulsion vector $\bm e$ to independently change in a diffusive manner in 3D or quasi-2D systems reduces the lateral swim pressure when $\bm e$ is oriented normal to the confining wall, i.e., the driving force for  cluster formation is reduced. Interestingly, in our systems the swim direction of the squirmers is preferentially parallel to the surfaces, as shown in Fig.~\ref{fig:prop_orientation}. Hence, despite a preferential 2D swimming motion, squirmers are less likely to form clusters in the quasi-2D geometry compared to the strict 2D case. Moderate fluctuations in the propulsion direction seem to be sufficient to modify the onset of cluster formation.

\begin{figure}[t]
  \centering
    \includegraphics[width=\columnwidth]{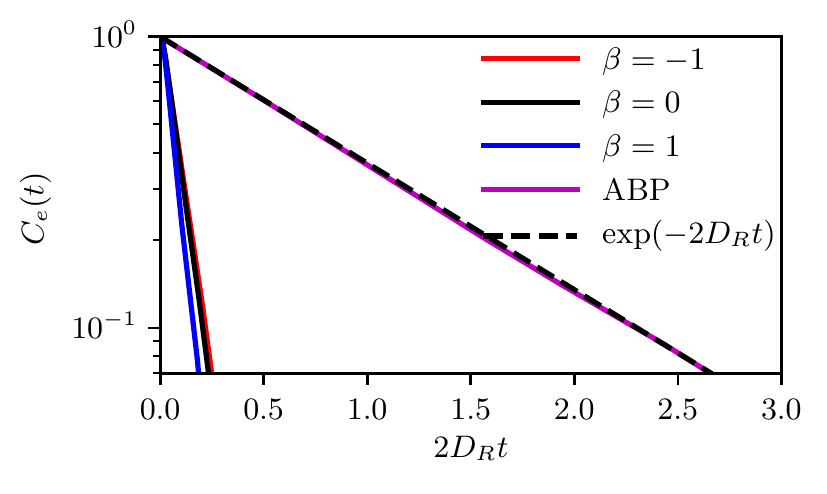}
  \caption{\label{fig:spherical_orientation}
  Orientation correlation function $C_e(t)$ averaged over all swimmers for a system of spherical squirmers and active Brownian particles at $\phi^{2D}=0.6$ and $Pe=115$.
  }
\end{figure}

The suppression of MIC and MIPS is attributed to a strong reorientation (scattering) of squirmers by hydrodynamic torques during their collisions \cite{mata:14}. This conclusion is quantitatively confirmed by our simulations in  Fig.~\ref{fig:spherical_orientation}, where the orientation autocorrelation function
\begin{align} \label{Eq:av_or_corr}
C_e(t)=\frac{1}{N_{sw}} \sum_{i=1}^{N_{sw}} \langle \bm{e}_i(t) \cdot \bm{e}_i(0) \rangle
\end{align}
for squirmers and ABPs is shown. It decays exponentially as $\exp(-t/\tau_{sq})$ with a characteristic time $\tau_{sq}$.  For ABPs, the relaxation time is given by $1/2D_R$, in agreement with theoretical expectations.  However, for squirmers, $\tau_{sq}$ is approximately an order of magnitude smaller, and rather similar for the various force-dipole strengths. This can be interpreted as a reduced effective P\'eclet number $Pe'=\tau_{sq} U_0/ R \approx 10$, a value, where no phase transition is expected even for ABPs. However, the differences in the tendency of the squirmers to form clusters (cf.~Fig.~\ref{fig:MIPS_suppressed_spherical}(a)) indicate the importance of hydrodynamic interactions, which are not captured in the P\'eclet number calculate on the basis of the rotational diffusion coefficient at infinite dilution.

\begin{figure}
  \centering
    \includegraphics[width=\columnwidth]{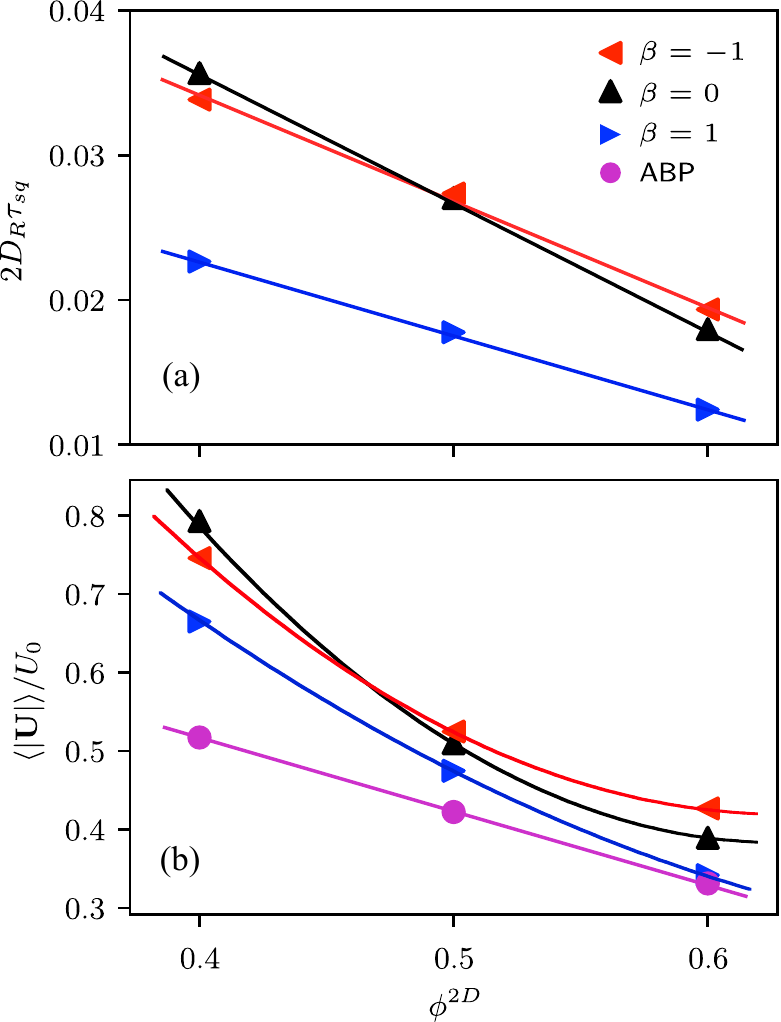}
  \caption{\label{fig:spherical_tausq}
 (a) Decay time $\tau_{sq}$ of the orientational correlation function $C_e(t)$ and (b) average swimming velocity as function of the squirmer concentration for neutral squirmers ($\beta=0$), pushers ($\beta=-1$), pullers ($\beta=1$), and active Brownian particles.  The decay time is scaled by the rotational relaxation time $1/2D_R$ and the swimming velocity by that of the set value $U_0$. The P\'eclet number is $Pe=575$. The lines are guides to the eye (either linear or parabolic functions).
  }
\end{figure}

The relaxation time $\tau_{sq}$ depends on the squirmer concentration, as shown in Fig.~\ref{fig:spherical_tausq}(a) for $Pe=575$. Compared to ABPs, $\tau_{sq}$ of squirmers is reduced by orders of magnitude (factor $50 - 80$) and decreases almost linearly in the considered range of concentrations (for $Pe=115$,  $2D_R \tau_{sq} \approx 0.1$ (cf. Fig.~\ref{fig:spherical_orientation}).) Naturally, $\tau_{sq}$ will increase in a nonlinear manner for $\phi^{2D} \to 0$, because at $\phi^{2D}=0$, $2 D_R \tau_{sq} =1$. In addition, the average swimming velocity $\lla |\bm U| \rra$ decreases with increasing concentration, also for ABPs (cf. Fig.~\ref{fig:spherical_tausq}(b)). Thereby, the velocity for ABPs seems to decrease linearly, whereas those of the various squirmers decline nonlinearly. A decreasing swimming velocity and relaxation time with increasing areal fraction have also been found in Ref.~\cite{mata:14} for strictly 2D systems.

\section{Simulations---Spheroidal Squirmers} \label{sec:results_spheroidal}

\begin{figure*}
  \centering
    \includegraphics[width=0.475 \columnwidth]{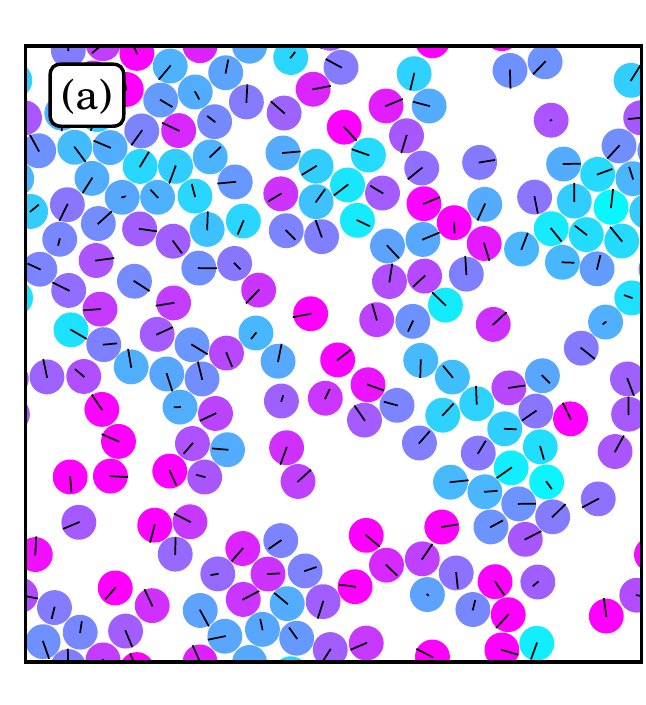}
    \includegraphics[width=0.475 \columnwidth]{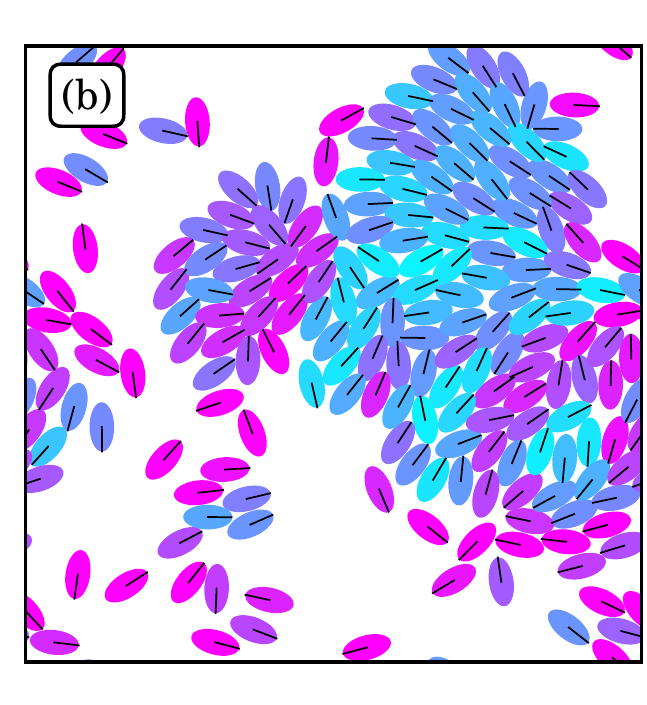}
    \includegraphics[width=0.475 \columnwidth]{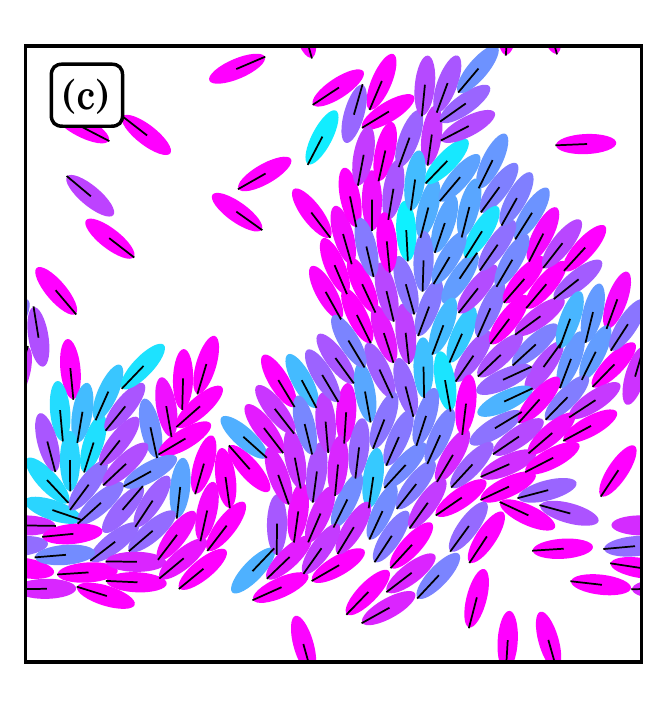}
    \includegraphics[width=0.6075 \columnwidth]{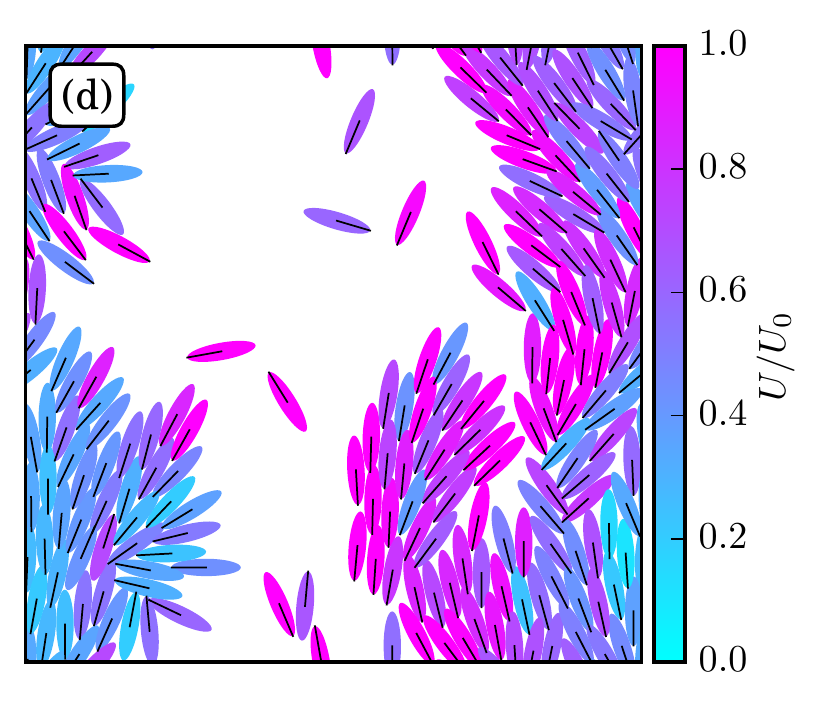}
  \caption{\label{fig:snapshots_aspect_1_to_4}
  Snapshots of the structure of neutral ($\beta=0$) spheroidal squirmers for $Pe=12$, the area packing fraction $\phi^{2D}=0.5$, and the aspect ratios (a) $b_z/b_x =1$,  (b) 2, (c) 3,  and (d) 4.  Similarly, the clusters in (b) are unstable. See Supplemental Material at ??? for movies of the various cases (M3 -- M6).}
\end{figure*}

\subsection{System Setup and Parameters}

We consider spheroidal squirmers with the minor axis $b_x=3a$ and the aspect ratios $b_z/b_x \in \{1,1.5,2,3,4\}$ in the quasi-2D geometry described above (see Fig. \ref{fig:slit_sketch}).   We employ the time step $h=0.02 \sqrt{ma^2/(k_BT)}$, rotation angle $\alpha=130^\circ$, and the mean number of particles $\lla N_c \rra=10$, which yields the fluid viscosity $\eta=17.8 \sqrt{m k_B T/a^4}$.

To investigate the collective behavior for different packing fractions $\phi= 4 \pi N_{sw} b_x^2b_z/3 L^2 L_y$, or equivalently area packing fractions
$\phi^{2D}=N_{sw} \pi b_x b_z/L^2$, the length $L$ of the simulation box is varied, while the number of swimmers is fixed at $N_{sw}=196$ for $b_z/b_x=1$ and $N_{sw}=200$ for all other aspect ratios.
We choose the values of the  swimming mode $B_1/\sqrt{k_BT/m}=0.01,0.007,0.005,0.003,$ and $0.002$, which yields, with Eq.~(\ref{eq:u0}), the identical P\'{e}clet number
\begin{align}
  Pe=U_0/(2D_R^\perp b_z)=12
\end{align}
for all aspect ratios. Here, $D_R^\perp=k_BT/\xi^\perp$ is the rotational diffusion coefficient around the minor axis. Note that this P\'eclet number is much smaller than those of the previous section for our studies on MIPS of spheres. As before, $\beta \in \{ -1,0,1\}$ and simulations of ABPs are performed for comparison.

\subsection{Results}

\begin{figure}
\includegraphics*[width=\columnwidth]{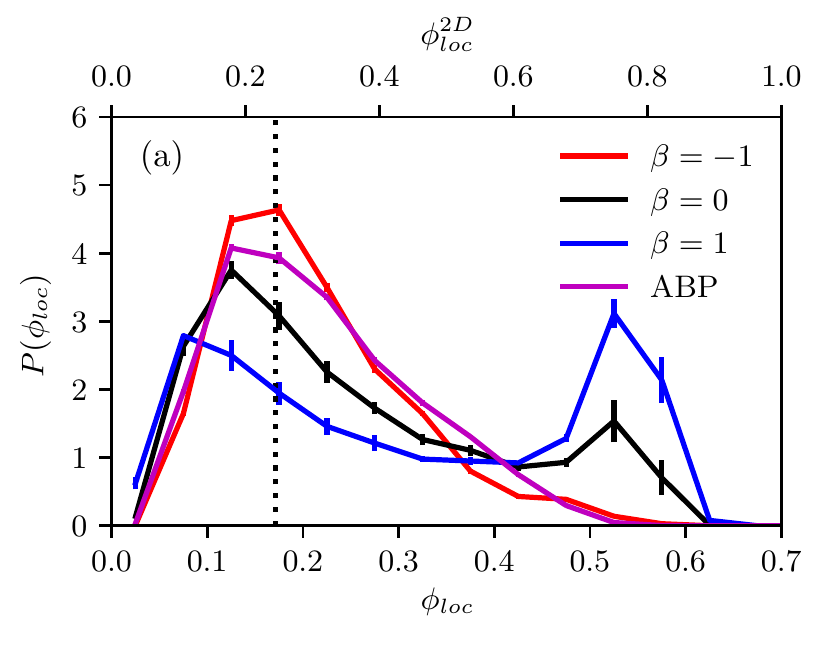} \\[5pt]
\includegraphics*[width=\columnwidth]{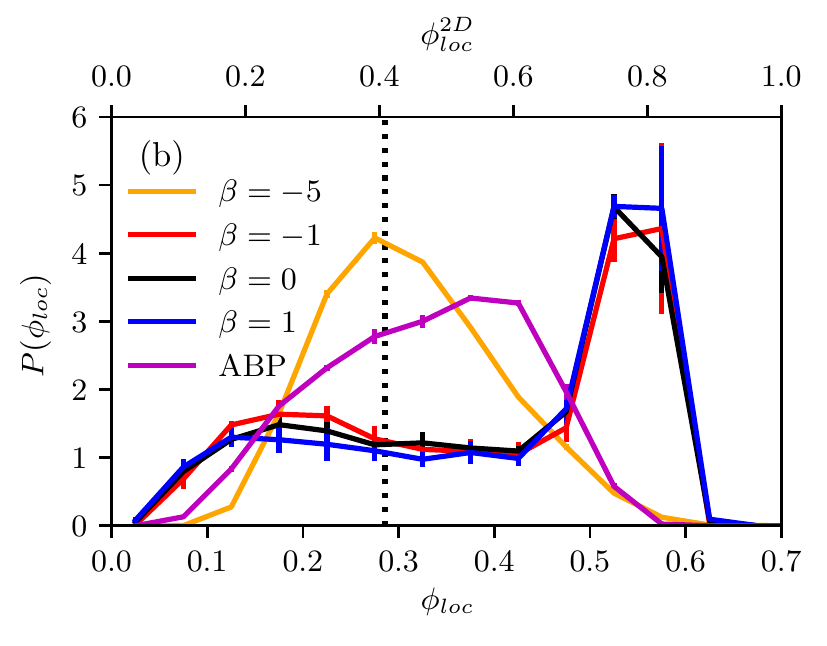}
\caption{\label{fig:phi50_a6_betas_ABP}
  (a) Probability distribution of the local packing fraction for spheroidal squirmers with the aspect ratio $b_z/b_x=2$ at the average area packing fraction $\phi^{2D}=0.3$, i.e.,  $\phi=0.17$. (b) Corresponding results for $\phi^{2D}=0.5$, i.e., $\phi=0.29$. The P\'eclet number is $Pe=12$.
}
\end{figure}

Figure~\ref{fig:snapshots_aspect_1_to_4} illustrates structure formation for various aspect ratios. Although the P\'eclet number $Pe = 12$ is quite small, spheroidal squirmers clearly reveal clusters for all aspect ratios $b_x/b_z \geqslant 2$. Thereby, a larger aspect ratio is beneficial for cluster formation. For spherical squirmers, no phase separation occurs at this P\'eclet number (Fig. \ref{fig:snapshots_aspect_1_to_4} (a)). The anisotropic nature of the spheroids leads to shape-induced jamming and alignment, where alignment increases with increasing aspect ratio as is evident from the comparison of Figs.~\ref{fig:snapshots_aspect_1_to_4}(b) and \ref{fig:snapshots_aspect_1_to_4}(d).

Again, we employ Voronoi tessellation to quantify clustering and phase separation. However, we cannot use  spheroid centers as generator points. For spheroids, the distance from a point in space to a swimmer is measured as the Euclidian distance to the nearest point on the swimmer's surface. For an efficient calculation of the (approximate) Voronoi volume, we employ a vertex-mesh on the surface of every spheroid \cite{pers:04}.
Those vertices serve as generator points and the Voronoi cell of a swimmer is defined as the union of the cells of its vertices.

The probability distribution $P(\phi_{loc})$ of the local packing fraction is displayed in Fig.~\ref{fig:phi50_a6_betas_ABP}. No phase separation occurs for ABPs and pushers ($\beta=-1$) at the concentration $\phi_{loc}^{2D}=0.3$ (Fig.~\ref{fig:phi50_a6_betas_ABP}(a));  however, clusters appear for neutral squirmers ($\beta=0$) and pullers ($\beta=1$). At the higher concentration $\phi_{loc}^{2D}=0.5$ (Fig.~\ref{fig:phi50_a6_betas_ABP}(b)), all squirmers with $\beta =0, \pm 1$ exhibit dense clusters.  In contrast, spheroidal APBs exhibit only a weak tendency to form clusters. This points toward a major role of hydrodynamics in cluster formation of elongated squirmers. Simulations of pushers with a strong active stress ($\beta= -5$) emphasized this aspect. Here,  we find a clear maximum in $P(\phi_{loc})$ at the average density and, hence, no MIPS.

\begin{figure}
\includegraphics*[width=\columnwidth]{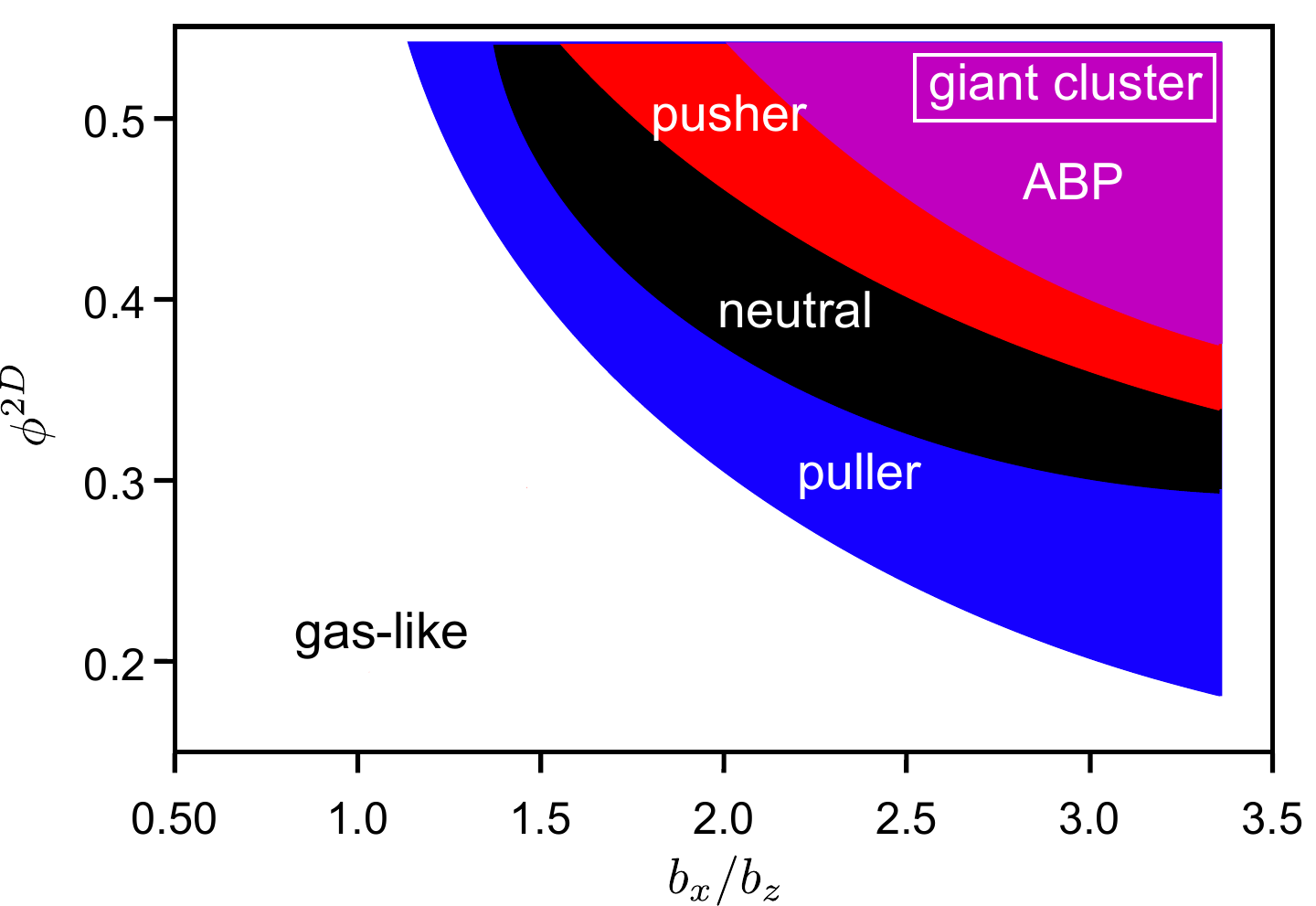}
\caption{\label{fig:state_diagram}
  State diagram for squirmers and ABPs at $Pe=12$. The blue, black, red, and purple areas indicate  giant clusters for pullers ($\beta=1$), neutrals ($\beta=0$), pushers ($\beta=1$), and active Brownian particles, respectively. Thereby clusters of pullers appear in all colored areas, clusters of neutral squirmers in the black, red, and purple area, etc. See Supplemental Material at ??? for individual state diagrams of the various swimmers. }
\end{figure}

By varying the aspect ratio ($b_z/b_x  \in \{1,1.5,2,3\}$) and the squirmer concentration ($\phi^{2D} \in \{0.1,0.2,0.3,0.4,0.5 \}$), we establish the state diagram displayed in Fig.~\ref{fig:state_diagram}. The various colored areas indicate the coexistence of dense cluster region and a dilute region. The respective phase borders should be viewed as approximations to the infinite-systems-size phase diagram. An accurate calculation of the phase boundaries requires a substantial simulation effort due to finite-size effects. The considered system sizes are certainly too small to ensure absence of finite-size effects.
The blue area indicates that pullers ($\beta=1$) already phase separate for relatively small aspect ratios and at low packing fractions. For decreasing $\beta$ (black area $\beta=0$,  red area $\beta=-1$), the phase-separation line shifts to the top right, i.e,  higher aspect ratios and packing fractions are required for phase separation to occur. The fact that the area for phase separation of ABPs is farthest to the top-right indicates that hydrodynamics enhances phase separation for elongated squirmers. Hence, the influence of hydrodynamic interactions is reversed compared to spherical swimmers, where hydrodynamics suppresses cluster formation. However, the qualitative effect of the active stress remains unchanged---pulling enhances cluster formation compared to pushing.



\section{Discussion} \label{sec:discussion}

Our simulation results for spherical squirmers agree with certain aspects of some previous studies, but disagree with others.  In particular, the  cluster-size  distribution function has been analyzed  in Ref.~\cite{alar:17} for the same squirmer model with the Lattice Boltzmann approach, but a somewhat different simulation setup.  Significantly larger systems with many more squirmers have been considered, and three-dimensional hydrodynamic interactions, but squirmers confined in 2D at the density $\phi^{2D} =  0.1$. We like to emphasize that our studies yield probability distribution functions for the local packing fraction of squirmers, which are independent of the system size. In agreement with our Fig.~\ref{fig:cluster_size}, in Ref.~\cite{alar:17} a power-law decay ($p(n) \sim n^{-2.0}$) for small cluster sizes ($n\lesssim 10^3$) is obtained for pullers with $0< \beta \lesssim 1$.  For other $\beta$ ($\beta >2$ and $\beta< 0$), a stronger decay appears for $n \gtrsim 10$ in \cite{alar:17}. This non-power-law decay is stronger than that observed in our simulations. Hence, at smaller concentrations we expect a stronger drop of the cluster distribution for $\beta <0$ also in our quasi-2D systems.

Our simulation results of spherical squirmers disagree qualitatively and quantitatively  with results of Refs.~\cite{zoet:14,blas:16}, where comparable systems have been considered. We find that hydrodynamic interactions suppress cluster formation of squirmers, in contrast to Ref.~\cite{zoet:14}. According to Ref.~\cite{blas:16}, the P\'eclet numbers of our neutral squirmers are deep in the phase-separated region of the phase diagram (Fig. 3 of Ref.~\cite{blas:16}) \cite{comm2}. However, we do not see any signature for phase separation, not even for neutral squirmer. Moreover, the distribution function of the bond order parameter differs significantly. For squirmers, we find the most pronounced local order for pullers, followed by neutral squirmers and weakest order for pushers, in agreement with other studies \cite{delm:15,yosh:17,alar:17}. The studies of Ref.~\cite{zoet:14} show strongest cluster formation and local order for neutral squirmers, comparable to strictly 2D ABPs and more pronounced compared to ABPs in quasi-2D.

We suspect that the origin of the discrepancy between our results and those of Refs.~\cite{zoet:14,blas:16} is the different compressibilities of the employed MPC fluids. It is shown in App.~\ref{app:inhomogeneities}  that the squirmer-induced fluid transport can cause strong MPC-fluid density inhomogeneities when squirmers are blocked, either by other squirmers or by a surface, at too strong propulsion velocities. A significant fluid-density modulation leads to a reduced propulsion efficiency and a reduced active pressure inside the cluster. As a consequence, propulsion of squirmers confined inside a cluster is effectively reduced, which leads to more stable clusters. By our choice of simulation parameters, we ensured  that critical fluid density variations are avoided. Thus, the results of the simulation studies of Refs.~\cite{zoet:14,blas:16} are incompatible with ours; however, if (strong) compressibility effects would be considered as relevant, like fluids near a critical point, these studies are applicable.

In the case of spheroidal squirmers,  two aspects contribute to the hydrodynamic enhancement of cluster formation, namely near-field HIs between squirmers, and between a squirmer and the confining walls \cite{thee:16.1}. The effect of the active stress $\beta$ is evident from the state diagram of Fig.~\ref{fig:state_diagram}. Note that the force (FD) and source dipole (SD) flow-field contributions of a squirmer \cite{thee:16.1} with $e_y =0$ decay as $\varrho^{-3}$  and $\varrho^{-2}$ in the center between the walls, respectively, with the distance $\varrho$ from a squirmer center parallel to the walls \cite{math:16.1}.
 Our simulation studies of Ref.~\cite{thee:16.1} on the cooperative motion of two squirmers in a slit geometry already reveal a long-time stable configuration of two pullers, in which they swim together in a wedge-like conformation with a constant small angle. A similar configuration of neutral squirmers or pushers is far less stable and the squirmers scatter more strongly, which emphasizes the relevance of specificities in fluid interactions of the squirmers due to their close proximity, and particular with the walls. Applying three-dimensional hydrodynamic interactions rather than no-slip boundary conditions to a still confined pair of pullers yields short cooperative motion only, emphasizing that surface hydrodynamics is extremely important \cite{thee:16.1}. Hence, the seed for the cluster formation is the cooperative motion of two squirmers. Blockage by additional encountering squirmers further stabilizes the emerging cluster. The latter is reflected in the density dependence of the cluster formation.

\section{Summary and Conclusions} \label{sec:conclusion}

We have studied the collective dynamics of spherical and spheroidal squirmers confined in a narrow slit by mesoscale hydrodynamic  simulations (MPC). To elucidate the role of hydrodynamic interactions, we have complemented the studies by  simulations of spherical and spheroidal active Brownian particles.

We find that hydrodynamics suppresses phase separation for spherical microswimmers, in contrast to Ref.~\cite{zoet:14}, but in accordance with Ref.~\cite{mata:14}. We attribute the contradiction with Ref.~\cite{zoet:14} to compressibility effects of the MPC fluid.  In agreement with  Ref.~\cite{mata:14}, we confirm that a hydrodynamically enhanced reorientation of squirmers during  swimmer-swimmer collisions implies an effectively smaller P\'eclet number and, hence, a suppression of phase separation. However, the squirmers form short-lived clusters over a broad range of cluster sizes. As already pointed out in Ref.~\cite{yosh:17}, near-field hydrodynamic interactions play a major role in cluster formation. This is reflected in the clear differences between pullers, neutral squirmers, and pushers \cite{yosh:17}. A strong near-field effect can be expected, since the multipole contributions to a swimmer's flow field decay quickly with distance parallel to the confining walls in a narrow channel \cite{math:16.1}.

For anisotropic swimmers, alignment due to steric interactions is a key mechanism for MIC, while isotropic swimmers form clusters due to jamming and blocking. Our studies of elongated, prolate spheroidal squirmers yield hydrodynamically  enhanced MIC compared to ABPs. This result is surprising at first glance. Based on our studies on the cooperative motion of pairs of confined elongated squirmers, we attribute this enhancement to near-field hydrodynamics and, most importantly, swimmer-surface hydrodynamic interactions. Yet, the anisotropic shape leads to shape-induced phase separation for ABPs at considerably smaller P\'eclet numbers than for spheres. This is related to differences in the clustering  mechanisms of isotropic and anisotropic swimmers.

For both, spherical and spheroidal squirmers, the active stress parameter $\beta$ correlates positively with cluster formation, i.e., pullers ($\beta>0$) show the strongest tendency to form clusters, followed by neutral squirmer ($\beta=0$), and pushers ($\beta<0$). In fact, this is well consistent with our previous study of two squirmers in a narrow slit, which shows that cooperativity is most pronounced for pullers \cite{thee:16.1}.

Simulations of dumbbell swimmers \cite{suma:14.1,sieb:17} in three dimensions have also been performed and  hydrodynamically enhanced MIC has been observed \cite{furu:14}. Since dumbbell swimmers are anisotropic, this finding is consitent with our results for spheroidal squirmers.
However, we have to be careful with too general conclusions. As we demonstrated, MIC depends decisively on the hydrodynamic flow field, i.e., the details of the flow field matters. This is reflected by our results for very strong pushers ($\beta=-5$), which show a suppression of phase separation.

The observed clustering of spheroidal squirmers is in contrast to the behavior of motile bacteria, e.g., {\em E. coli}, in suspensions, which display active turbulence \cite{wens:12}.  At the moment, there is no detailed understanding of the different behaviors and underlying mechanisms. Could a dependence on the P\'eclet number explain the differences? The rotational diffusion coefficient $D_R^e=0.057/s$ of non-tumbling {\em E. coli}  has been measured in Ref.~\cite{wens:12}. With the swimming velocity $U_0^e=30 \mu/s$ and the length $2b_z\approx 10 \mu$, we obtain a P\'eclet number $Pe^e \approx 50$. Our simulations show that such  $Pe$ even further enhance clustering. There seems to be an additional relevant interaction in bacteria suspensions. Of course, it could be related to the structure of bacteria with the cell body and the flagella bundle \cite{berg:04}, which form a rather flexible structure. Depending on the body-bundle orientation, a bacterium exhibits a wobbling motion, which perturbs alignment and leads to a fast decay of orientational correlations of nearby swimmers \cite{eise:16.1}. Furthermore, hydrodynamic interactions could be responsible---the propulsion of the bacterium with a rotating flagella bundle and a counterrotating cell body leads to a rotlet dipole \cite{hu:15}. Such a flow field leads to circular motion on a planar surface \cite{hu:15,laug:06,dile:11,leme:13} and may also enhance the decorrelation of the orientational order.

Squirmers are highly idealized models of biological microswimmers, not only in terms of shape, but also in terms of the stationary, time-independent flow fields. The strong interplay between shape and hydrodynamics, which is revealed by our simulations, indicates that the details of the cell shape and beat pattern may play a much more important role in the collective dynamics of biological microswimmers than previously expected.

\begin{acknowledgments}
We thank A. Wysocki and R. Hornung for helpful discussions. Support of this work by the DFG priority program SPP 1726 on ``Microswimmers -- from Single Particle Motion to Collective Behaviour'' is gratefully acknowledged.
We also acknowledge partial financial support by the Bavarian Ministry of Economic Affairs, Energy and Technology within joint projects between Helmholtz institutes.
In addition, the authors gratefully acknowledge the computing time granted through JARA-HPC on the supercomputer JURECA at Forschungszentrum J\"ulich.
\end{acknowledgments}

\appendix

\section{Simulation Approaches} \label{app:simulation}

\subsection{Multiparticle Collision Dynamics (MPC)} \label{app:mpc}

In MPC \cite{male:99,kapr:08,gomp:09}, a fluid is represented by $N$ point particles of mass $m$ with continuous positions $\bm{r}_i$ and velocities $\bm{v}_i$ ($i\in \{1,\ldots,N\}$).
The particle dynamics proceeds in two steps---streaming and collision.
During the streaming step of duration $h$ (collision time), the particles move ballistically, i.e., their positions are updated according to
\begin{align} \label{EqSolventStreaming}
 {\bm r}_i(t+h)={\bm r}_i(t)+ h {\bm v}_i(t).
\end{align}
In the collision step, the particles interact locally by an instantaneous stochastic process, for which we apply the stochastic rotation dynamics (SRD)  variant of the MPC method \cite{male:99,kapr:08,gomp:09,ihle:01} with angular momentum conservation (MPC-SRD+a) \cite{nogu:08,yang:15,thee:15}. For this purpose, the simulation volume is partitioned into cubic collision cells of length $a$.
In each of the cells the  particle velocities are then updated as \cite{nogu:08,thee:15}
\begin{align}
\begin{split}
 {\bm v}_i^{new} = & ~{\bm v}_{cm}+\mathrm{\bf R}(\alpha){\bm v}_{i,c}   \\
	  & - {\bm r}_{i,c} \times \Big[ m \mathrm{\bf I}^{-1} \sum_{j \in cell}\left\{{\bm r}_{j,c}\times \left({\bm v}_{j,c}-\mathrm{\bf R}(\alpha) {\bm v}_{j,c} \right) \right\} \Big].
\end{split}
\end{align}
Here, $\mathrm{\bf R}(\alpha)$ is the rotation matrix for the rotation of the relative velocity ${\bm v}_{i,c}={\bm v}_{i}-{\bm v}_{cm}$, with respect to center-of-mass velocity $\bm{v}_{cm}$ of the cell, by an  angle $\alpha$ around a randomly oriented axis.  ${\bm r}_{i,c}={\bm r}_{i}-{\bm r}_{cm}$ is the position relative
to the center of mass of the cell ${\bm r}_{cm}$, and $\mathrm{\bf I}$ is the moment of inertia tensor of the particles in the cell's center-of-mass frame.
The grid of collision cells is shifted randomly before each collision step to ensure Galilean invariance \cite{ihle:03}. Thermal fluctuations are intrinsic to the MPC method \cite{huan:10,huan:15,thee:15}. Hence,
a cell-level canonical thermostat (MBS thermostat) is applied after every collision step, which maintains the  temperature $T$ \cite{huan:10,huan:15}.
The algorithm conserves mass, linear, and angular momentum on the collision cell level, which gives rise to hydrodynamics on large length and long time scales \cite{kapr:08,ihle:09,huan:12}.

We measure lengths in units of the collision cell size $a$ and time units of $\sqrt{ma^2/k_BT}$, where $k_B$ is the Boltzmann factor, which corresponds to the choice $a=m=k_BT=1$. The rotation angle is set to $\alpha=130^\circ$. The viscosity $\eta$ of the fluid can be tuned by changing the mean number of particles in a cell $\lla {N}_c \rra$ or the time step $h$ \cite{nogu:08,thee:15,huan:15}.

\subsection{Boundary Conditions and Squirmer Implementation} \label{app:bc}

No-slip boundary conditions are implemented at the confining walls (cf. Fig.~\ref{fig:slit_sketch})  via the bounce-back rule $\bm{v}_i'(t)=-\bm{v}_i(t)$, where $\bm{v}_i$ and $\bm{v}_i'$ are the velocity before and after the particle's collision with the wall, respectively. Additionally, spatially uniformly distributed phantom particles with Gaussian distributed velocity components are inserted into the walls and interact with fluid particles during the collision step \cite{lamu:01,gomp:09}. Parallel to the walls, periodic boundary conditions are applied.

During the streaming step, a squirmer moves like a passive colloid according to rigid-body equations of motion. The equations of motion of the rotational degrees of freedom are solve by introducing quaternions \cite{thee:16.1,alle:87} (see also App.~\ref{app:abp}). Interacting fluid particles experience the modified bounced back rule $\bm{v}_i'=-\bm{v}_i+2\bm{v}_S$, where $\bm{v}_S$ is the squirmer's surface velocity at the point of contact. This surface velocity includes its translational and rotational velocity as well as its slip velocity $\bm u_{sq}$ (\ref{Eq:Def_usq_of_zeta}). The change of a fluid particle's linear and angular momentum is transferred to the squirmer, such that the total momenta are conserved.
For the collision step, as for walls, spatially uniformly distributed phantom particles are generated inside the squirmer with Gaussian distributed velocity components, where their mean value is given by the rigid-body translational and rotational velocity plus the squirming velocity, and the variance is equal to $k_BT/m$ \cite{thee:16.1,goet:10}. After the collision step, the linear and angular momentum changes of all phantom particles are transferred to the squirmer. A more detailed description is provided in Ref. \cite{thee:16.1}.
Repulsive forces and torques due to steric interactions between squirmers, and between a squirmer and a wall, during a streaming step are implemented as outlined in the appendix of Ref. \cite{thee:16.1} (see also Ref. \cite{para:05}).

\subsection{Simulation Algorithm for Spheroidal Active Brownian Particles} \label{app:abp}

For active Brownian particles, as for the hydrodynamic case, we solve the rigid-body equations of motion, but now in the presence of translational and rotational noise. Thereby,  the squirmer orientation is parameterized by a unit quaternion $\bm{q}=(q_0,q_1,q_2,q_3)$ \cite{alle:87}. The integration scheme for the overdamped dynamics of the  center-of-mass $\bm R(t)$ and the quaternion $\bm{q}(t)$ of the spheroid is \cite{ilie:15}
\begin{align}
  \bm{R}(t+h)&=\bm{R}(t)+\bm{\mu} \bm{F} h +U_0\bm{e}h, \\ \label{eq:quat}
  \bm{q}(t+h)&=\bm{q}(t)+\frac{1}{2}\bm{Q}(\bm{q}(t))  \begin{pmatrix}   0 \\ \bm{D} \bm{\mu}^R \bm{T}  \end{pmatrix} +\lambda_q \bm{q}(t),
\end{align}
where $h$ is the time step, $\bm F = \bm F^{st} + \bm F^{th}$ and  $\bm{T}=\bm{T}^{st}+\bm{T}^{th}$ are the force and torque on the ABP due to steric interactions and thermal noise. The $4\times4$ matrix $\bm Q$ comprised of the quaternions is defined in Appendix~\ref{app:SpheroidalABP}. As before, $U_0$ is the propulsion velocity and $\bm{e}\equiv \bm{D}^T \bm{e}_z$ points  along the major axis of the squirmer  and in the direction of the desired propulsion. The rotation matrix $\bm{D}$ that rotates vectors from the laboratory frame into the body-fixed frame is given in Appendix~\ref{app:SpheroidalABP}. The Lagrangian multiplier $\lambda_q$ is introduced to ensure normalization of $\bm{q}$. Hence, we perform the quaternion update without the term $\lambda_q \bm{q}(t)$ to obtain some $\tilde{\bm{q}}$ and then determine $\lambda_q$ such that $\tilde{\bm{q}}+\lambda_q \bm{q}$ is normalized \cite{ilie:15}.

The mobility matrix $\bm{\mu}$ relates the applied force $\bm{F}$ to the resulting ABP velocity $\bm{U}$ via $\bm{U}=\bm{\mu}\bm{F}$, while the rotational
mobility matrix $\bm{\mu}^R$ relates the applied torque $\bm{T}$ to the resulting angular velocity $\bm{\Omega}$ via $\bm{\Omega}=\bm{\mu}^R\bm{T}$.
In the body-fixed frame, the  respective mobility matrices are diagonal, with diagonal elements given by the inverse translational and rotational friction coefficients $\gamma_x^{-1},\gamma_y^{-1},\gamma_z^{-1}$ and $\xi_x^{-1},\xi_y^{-1},\xi_z^{-1}$, respectively. For a prolate spheroid $\gamma_x=\gamma_y=\gamma^\perp$, $\gamma_z=\gamma^\parallel$, $\xi_x=\xi_y=\xi^\perp$, $\xi_z=\xi^\parallel$. The parallel and perpendicular friction coefficients are provided in Appendix~\ref{app:SpheroidalABP}.  For zero eccentricity, i.e., a sphere with $b_x=b_z=R$, we obtain $\gamma^\parallel=\gamma^\perp=6 \pi \eta R$ and $\xi^\parallel=\xi^\perp=8 \pi \eta R^3$.

In the body-fixed frame, the Cartesian components of the thermal force $\bm{F}^{th}$ are Gaussian distributed random numbers of zero mean and variance $2k_B T \gamma_\alpha/h$  ($\alpha \in \{x,y,z\}$). The thermal torque $\bm{T}^{th}$ is generated similarly, but with variance $2k_B T \xi_\alpha/h$. The implementation of the repulsive forces and torques due to steric interactions between ABPs and between an ABP and a wall is described in the appendix of Ref.~\cite{thee:16.1}.

\section{Rigid-Body Dynamics: Quaternion} \label{app:SpheroidalABP}

The rotation matrix $\bm D$ and the matrix $\bm Q$ introduced in Eq.~(\ref{eq:quat}) are given by
\begin{align*} \label{Eq:Rotation_matrix_for_RBD}
 & \bm D = \\
 & \begin{pmatrix}
          q_0^2+q_1^2-q_2^2-q_3^2 & 2(q_1 q_2+q_0 q_3) & 2(q_1 q_3-q_0 q_2) \\
          2(q_2q_1-q_0q_3) & q_0^2-q_1^2+q_2^2-q_3^2 & 2(q_2q_3+q_0q_1)\\
          2(q_2q_1+q_0q_2) & 2(q_3q_2-q_0q_1) & q_0^2-q_1^2-q_2^2+q_3^2
          \end{pmatrix} ,
\end{align*}
\begin{align}
  \bm{Q}(\bm{q})=\begin{pmatrix}  q_0 & -q_1 & -q_2 & -q_3 \\ q_1 & q_0 & -q_3 & q_2 \\ q_2 & q_3 & q_0 & -q_1 \\ q_3 & -q_2 & q_1 & q_0
  \end{pmatrix}
\end{align}
in terms of the rotation quaternion $\bm{q}=(q_0,q_1,q_2,q_3)$  \cite{alle:87}.

The translational and rotational friction coefficients of a spheroid are given by
\begin{align}
  \gamma^\parallel &= 6 \pi \eta b_z \frac{8}{3}\hat e^3\left(-2 \hat e+(1+\hat e^2)L\right)^{-1}, \label{Eq:gamma_parallel}\\
  \gamma^\perp &=6 \pi \eta b_z \frac{16}{3}\hat e^3\left(2\hat e+(3\hat e^2-1)L\right)^{-1}, \label{Eq:gamma_perp} \\
  \xi^\parallel&=8 \pi \eta b_z^3  \frac{4}{3}\hat e^3(1-\hat e^2)\left(2 \hat e-(1-\hat e^2)L\right)^{-1}, \label{Eq:xi_parallel}\\
  \xi^\perp&=8 \pi \eta b_z^3  \frac{4}{3}\hat e^3(2-\hat e^2)\left(-2\hat e+(1+\hat e^2)L\right)^{-1}, \label{Eq:xi_perp}
\end{align}
where $\eta$ is the fluid viscosity, $\hat e=1/\tau_0=c/b_z$ is the eccentricity, and $L=\log \left( (1+\hat e)/(1-\hat e) \right)$. For $\hat e=0$, i.e., a sphere with $b_x=b_z=R$, follows $\gamma^\parallel=\gamma^\perp=6 \pi \eta R$ and $\xi^\parallel=\xi^\perp=8 \pi \eta R^3$.

\section{Avoiding MPC Simulation Artifacts Related to Density Inhomogeneities} \label{app:inhomogeneities}

\begin{figure}
  \centering
    \includegraphics[width=0.9\linewidth]{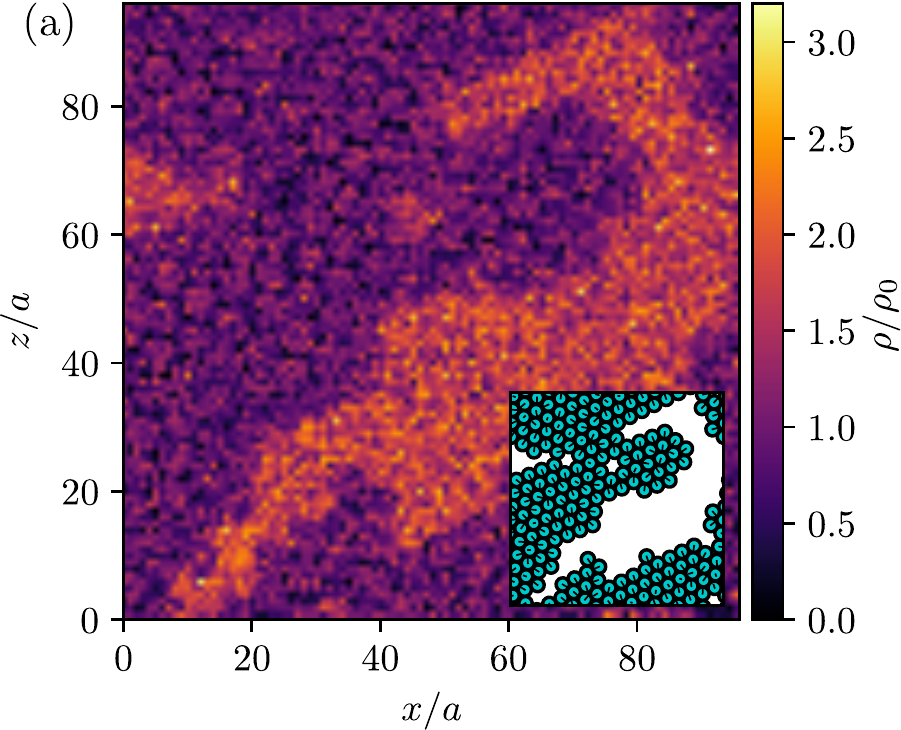} \\
    \includegraphics[width=0.9\linewidth]{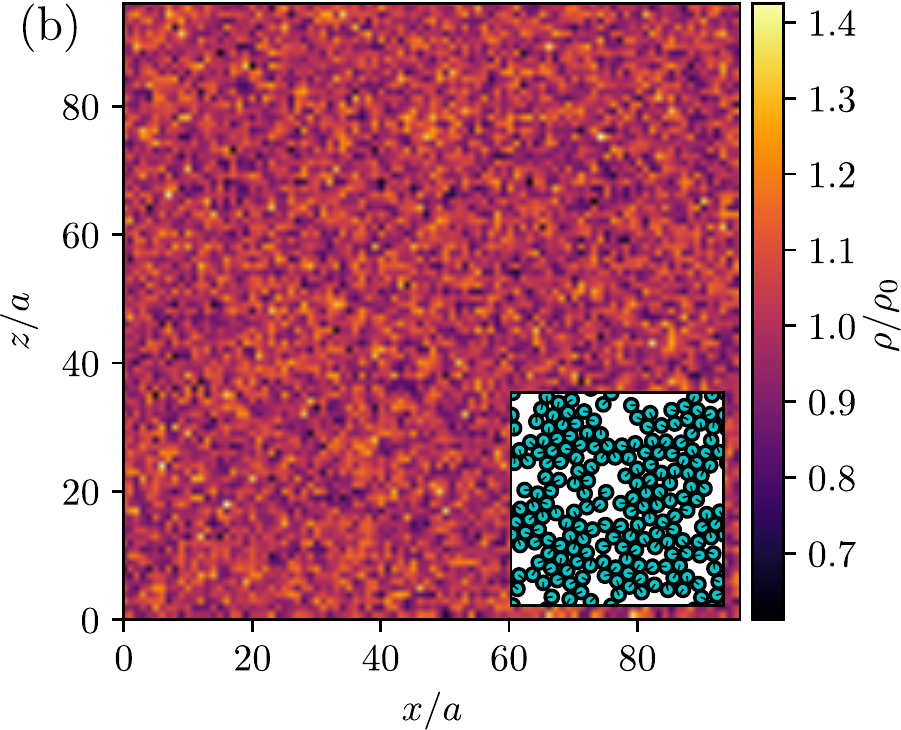}
  \caption{\label{fig:vacuum_pump}
  Fluid particle density  $\rho$ in a slit with spherical neutral squirmers at $Pe=115$ and $\phi^{2D}=0.6$. The total fluid density is show, comprising the MPC particle density $\rho_{fl}$ and that of the phantom particles $\rho_{ph}$, i.e., $\rho= m \langle N_c \rangle /a^3 = \rho_{fl}+\rho_{ph}$. $\rho_0$ denotes the average density. In (a), the choice $B_1=0.1 \sqrt{k_BT/m}$ and $\lla N_c \rra =10$ yields the pumping number $Pu=5$, and in (b), $B_1=0.01 \sqrt{k_BT/m}$ and $\lla N_c \rra =80$ yields $Pu=0.5$; the other parameters are the same. The insets show the corresponding squirmer locations. Please, note the differences in the color bars in (a) and (b).
  }
\end{figure}

MPC is a compressible fluid and correspondingly strong density inhomogeneities can emerge for certain choices of MPC parameters. Figure~\ref{fig:vacuum_pump} provides an example of the fluid-particle density distribution for neutral spherical squirmers confined in a narrow slit (cf. Fig.~\ref{fig:slit_sketch}).  The squirmer density is $\phi^{2D}=0.6$ and the P\'eclet number $Pe=115$. However, the P\'eclet number is realized by different combinations of MPC fluid and squirmer parameters. In Fig.~\ref{fig:vacuum_pump}(a), we set $B_1=0.1 \sqrt{k_BT/m}$ and $\lla N_c \rra =10$, and in (b) $B_1=0.01 \sqrt{k_BT/m}$ and $\lla N_c \rra =80$. The latter gives a ten times larger viscosity. The other parameters are the same as defined in App.~\ref{app:mpc} and \ref{sec:setup_low}.

Figure~\ref{fig:vacuum_pump}(a) shows large density inhomogeneities with dense (orange) and  rarefied regions (purple). Thereby, the rarefied regions coincides with the area occupied by the large squirmer cluster (cf. inset of Fig.~\ref{fig:vacuum_pump}(a)). In contrast, the MPC fluid distribution in Fig.~\ref{fig:vacuum_pump}(b) is rather homogeneous aside from thermal fluctuations and no link to the squirmer distribution is evident.
It seems that a pressure gradient is build up by active clustered squirmers by expelling fluid form the cluster---an effect related to the compressibility of MPC.
For density inhomogeneities as large as those in Fig. \ref{fig:vacuum_pump}(a), MPC simulations do not describe incompressible fluids, since the viscosity of the MPC fluid depends linearly on density and would, hence, depend strongly on position.

The mechanism in a MPC fluid, which opposes the diminished density in the gaps between squirmers is fluid-particle diffusion.  The ratio of the time $t_{diff}$ necessary for a MPC particle to diffuse over a swimmer diameter $\sigma=2R$, compared to the time required for a MPC particle to be advected by the surface activity of a squirmer with the velocity $U_0$ over the same distance, yields the dimensionless pumping number
\begin{align}
  Pu \equiv \frac{\sigma^2/(6D_f)}{\sigma/U_0}=\frac{\sigma U_0}{6 D_f},
\end{align}
where $D_f=0.013 \sqrt{a^2 k_BT}{m}$ is the MPC-fluid diffusion coefficient. For small pumping number, $Pu$, we expect the density inhomogeneities to disappear, i.e., MPC fluid-particle diffusion is faster than activity-induced advection.
The above choices of parameters yield the pumping numbers  $Pu=5$ (Fig.~\ref{fig:vacuum_pump}(a)) and $Pu=0.5$ (Fig.~\ref{fig:vacuum_pump}(b)).
Consistent with our expectation, the difference in the pumping number explains the appearing density modulations.

In terms of the P\'eclet number, $Pu$ is given by $Pu= \sigma^2 D_R Pe/6 D_f$. Since $Pe \gg 1$ for active systems, $Pu <1$ requires $\sigma^2 D_R/6D_f \ll 1$. Hence, the MPC parameters have to be chosen such that $D_R \ll 1$. Because $D_R \sim 1/\eta$, this is achieved for $\lla {N}_c \rra \gg 1$. Such an increase in  $\lla {N}_c \rra$ leaves $D_f$ virtually unchanged,  however,  implies an increase in computational effort.



\begin{thebibliography}{86}
\bibitem [{com()}]{comm2}%
  \BibitemOpen
  \href@noop {} {}\bibinfo {note} {Note that the definition of the P\'eclet
  number of Ref.~\cite{blas:16} is different; it is three times
  ours.}\BibitemShut {Stop}%
\end{thebibliography}

%

\end{document}